\documentclass{aa}
\usepackage{graphicx}

\begin{document}
\title{Are young radio sources in equipartition?}
%
\author{
M. Orienti\inst{1,2,3} \and
D. Dallacasa\inst{2,3} 
}
\offprints{M. Orienti}
\institute{
Instituto de Astrofisica de Canarias, 38200-La Laguna, Tenerife, Spain \and
Dipartimento di Astronomia, Universit\`a di Bologna, via Ranzani 1,
I-40127, Bologna, Italy \and 
Istituto di Radioastronomia -- INAF, via Gobetti 101, I-40129, Bologna,
Italy
}
\date{Received \today; accepted ?}

\abstract
{}
{The knowledge of physical conditions in young radio sources is
  important for defining the framework of models describing radio source
  evolution. We investigate whether young radio sources are in
  equipartition (i.e. minimum energy) conditions
by comparing the equipartition magnetic
  fields of 5 High Frequency Peakers (HFP) 
with values directly inferred from 
the spectral peak assumed to be produced by synchrotron self absorption.}
{Multi-frequency VLBA observations of 5 HFPs
  were carried out in both the optically thick and thin part of the
  spectrum to determine the spectral shape and
  angular size of the components for which individual
radio spectra were obtained.}  
{
  We find that the magnetic fields measured using observations agree
  well with those obtained by assuming equipartition, which implies
  that these sources are in minimum energy condition and the turnover
  in their spectra is due probably to SSA.
  In two source components, we found that
  the peak of the spectrum is caused by absorption of a thermal plasma
  instead of being due to SSA. The magnetic fields found in the various
  components range from 10 to
  100 mG. In the presence of such high magnetic fields, electron
  populations with rather low $\gamma$ emit in the GHz-regime.
  In one source, we detect low-surface brightness extended emission at
  low frequency located $\sim$ 30 mas ($\sim$ 50 pc)
from the main source. 
  This feature may be related to either an earlier episode of
  radio activity or a
  discontinuous start of the radio activity (sputtering).
  By comparing our data with previous VLBA observations, we
  estimate the hotspot advance speed to be in the range 0.1-0.7$c$ 
and kinematic ages of a few
  hundred years.}
{}
\titlerunning{Are young radio sources in equipartition?}
\maketitle
\section{Introduction}

In a scenario where radio sources grow in a self-similar way, the
evolution of each radio object originated by an Active Galactic 
Nucleus (AGN) depends on its linear size. It is 
established 
that the population of intrinsically compact radio sources represents
an early stage of radio source evolution. 
In such a context,
the empirical anti-correlation found between the linear
size and the turnover frequency (O'Dea \& Baum \cite{odea97}) implies
that the youngest sources - with ages of about $\sim$ 100 - 1000 years -
must be sought among those whose turnover
frequency occurs at a few GHz, which are termed ``High Frequency
Peakers'' (HFPs, Dallacasa et al. \cite{dd00}; Dallacasa \cite{dd03}).\\
Several evolution models (e.g. Fanti et al. \cite{cf95}; Readhead et
al. \cite{read96}; Snellen et al. \cite{sn00})  
were proposed to describe how the physical
parameters (i.e. luminosity, linear size and velocity) evolve as the
radio emission grows within the host galaxy. Such evolution models  
are based on the 
assumption of minimum energy content, corresponding to a near
equipartition of energy between the radiating particles and the
magnetic field (Pacholczyk \cite{pacho70}) usually referred to as
``equipartition''. 
However, there is no {\it a priori} reason for believing that magnetic
fields in radio sources are in equipartition.
It is thus important to test the reliability of the equipartition
at various stages of the radio source evolution;
the
study of newly born radio sources provides useful information
about the physical parameters at the beginning of
source growth.
If the spectral peak is due to pure synchrotron self-absorption
(SSA) as claimed by Snellen et al. (\cite{sn00}), 
we have an independent way of estimating
the magnetic field by measuring the frequency at which the radio emission
changes from being opaque to transparent. In this case, the magnetic field
$H$ can be measured directly from the spectral peak parameters,
namely the peak frequency $\nu_{p}$, peak flux density $S_{p}$, and
the source component 
angular sizes $\theta_{maj}$ and $\theta_{min}$ (Kellermann \&
Pauliny-Toth \cite{kpt81}):  

\begin{equation}
H \sim f(\alpha)^{-5} \theta_{maj}^{2} \theta_{min}^{2} \nu_{p}^{5}
S_{p}^{-2}(1+z)^{-1} 
\label{h_synchro}
\end{equation}

\noindent The function $f(\alpha)$ only weakly depends on $\alpha$,
and for $\alpha = 0.5$, $f(\alpha) \sim 8$ (Kellermann \& Pauliny-Toth
\cite{kpt81})\footnote{Throughout this paper, 
the radio spectrum is described by the power
  law $S \propto \nu^{- \alpha}$.}. The main difficulty in applying
this method has been the uncertainty in determining source
component parameters at the turnover frequency.
Since the magnetic field depends strongly on these observable
quantities, the estimates of
the field strength
derived so far are not very accurate. In particular, the uncertainties
in the component size, a parameter that may be difficult to
measure, contribute to the limited accuracy
of the magnetic
field estimates. Furthermore, magnetic fields
should not be measured for the entire source but for its
individual sub-structures, which are more likely to represent 
``homogeneous'' components such as those studied in analytic models.\\
Scott \& Readhead (\cite{sr77}) and Readhead (\cite{read94})
computed the magnetic field for sources of low-frequency spectral
turnovers close in value to the observing frequency 
and found that the magnetic fields inferred directly from the
spectrum were within a
factor of 16 of the equipartition values. However, there are no
systematic studies of sources with spectra peaking at higher
frequencies, i.e. objects younger than those in Scott \& Readhead
(\cite{sr77}). \\
Spangler et al. (\cite{spangler83}) studied the magnetic
field in the GHz-peaked spectrum (GPS) sources DA\,193 (J0555+3948) and
B1848+283, which were both classified as 
HFP by Dallacasa et al. (\cite{dd00}). 
They found that B1848+283 was in equipartition, while
DA\,193 was not. On the other hand, 
Orienti \& Dallacasa
(\cite{mo08a}) investigated the magnetic field in the very young GPS
object J1459+3337 and did not find any evidence of departure from
equipartition. It can be considered to be one of the youngest objects
known so far.\\
However, it is possible that the turnover in the spectrum is
caused by free-free absorption (FFA) in the external ionized medium, as
found for a number of young radio sources (e.g. Kameno et
al. \cite{kameno00}; Marr et al. \cite{marr01}). In this case, the
magnetic field inferred from the peak would be physically meaningless.\\ 
We determine the magnetic fields of the components of 5 High
Frequency Peaker radio sources from the Bright HFP sample
(Dallacasa et al. \cite{dd00}). The turnover frequency has a value of
a few GHz, corresponding to frequencies sampled well by VLBA
observations; this implies that the magnetic field strength of these
objects can be readily determined from spectral parameters and we can
infer whether such objects are in equipartition.

Throughout this paper, we assume the following cosmology: $H_{0} = 71\,
{\rm km\, s^{-1}\, Mpc^{-1}}$, $\Omega_{\rm M} = 0.27$ and
$\Omega_{\Lambda} = 0.73$, in a flat Universe.\\

\section{Target sources}

The ``bright'' HFP sample (Dallacasa et al. \cite{dd00}) consists of 55
candidate young radio sources 
selected on the basis of the their convex and high-frequency peaking
radio spectrum. However, 
not all the objects are genuinely young radio sources
but there is a significant fraction of contaminant
blazar objects that match the selection criteria during a particular phase of
their characteristic variability (e.g. Dallacasa et al. \cite{dd00}).\\
Several studies
were completed
to determine the real nature of each object. For the
majority of their lifetime, genuinely young radio sources and blazars
have different properties. Young radio sources usually do not
show significant spectral variability, and their radio emission - not
exceeding the pc-scale and
with a Double/Triple morphology dominated by lobes and/or
hot-spots - is generally unpolarized. On the
other hand, blazars are variable objects, and their emission,
with a core-jet structure, is strongly polarized.\\
For this reason, we selected five of the HFPs that turned out to be genuine
young radio sources on the basis of the 
results provided by investigations of spectral
variability (Orienti et al. \cite{mo07}; Tinti et al. \cite{st05}), morphology
(Orienti et al. \cite{mo06}), and polarization (Orienti \& Dallacasa
\cite{mo08b}). \\
The targets have a convex radio spectrum peaking at frequencies
between $\sim$ 4 and 15 GHz, and their pc-scale radio emission is
resolved into 2-3 well-separated 
individual sub-components, each with a radio
spectrum peaking at frequencies around a few GHz or higher.
Therefore, these sources are ideal targets to investigate the
physical conditions in young radio objects, as well as
the intrinsic magnetic field of each sub-component.\\
The observed objects are listed in Table \ref{parameter}. 
Together with
J1459+3337, they make up a small sample of 6 objects suitable for
testing whether 
equipartition magnetic field may apply even in the earliest phase
of the life of a radio source. The parameters for J1459+3337, 
listed in Table \ref{tab_mag}, are from Orienti \& Dallacasa
(\cite{mo08a}).\\ 

\section{VLBA observations and data reduction}

\begin{figure*}
\begin{center}
\includegraphics{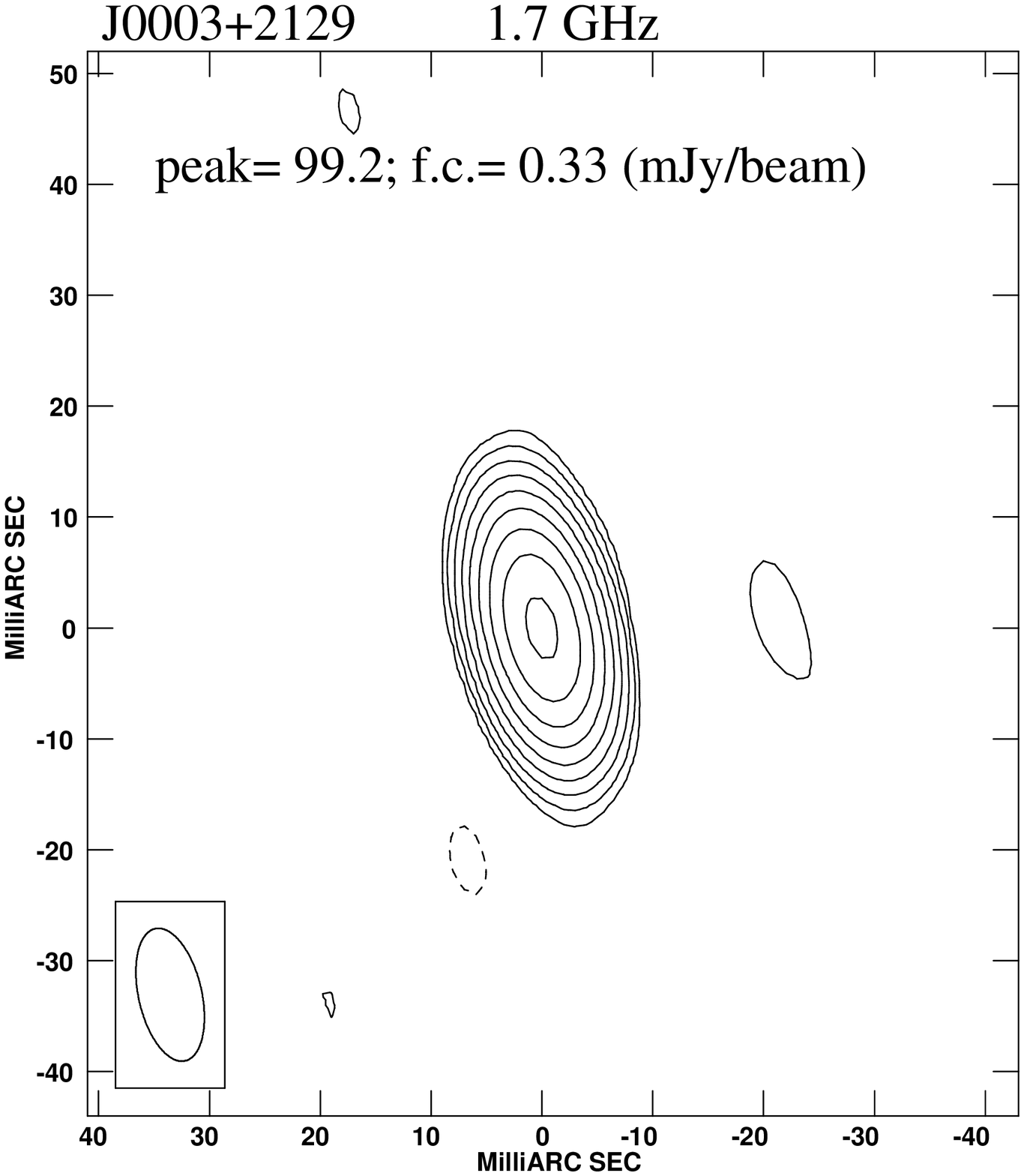}
\includegraphics{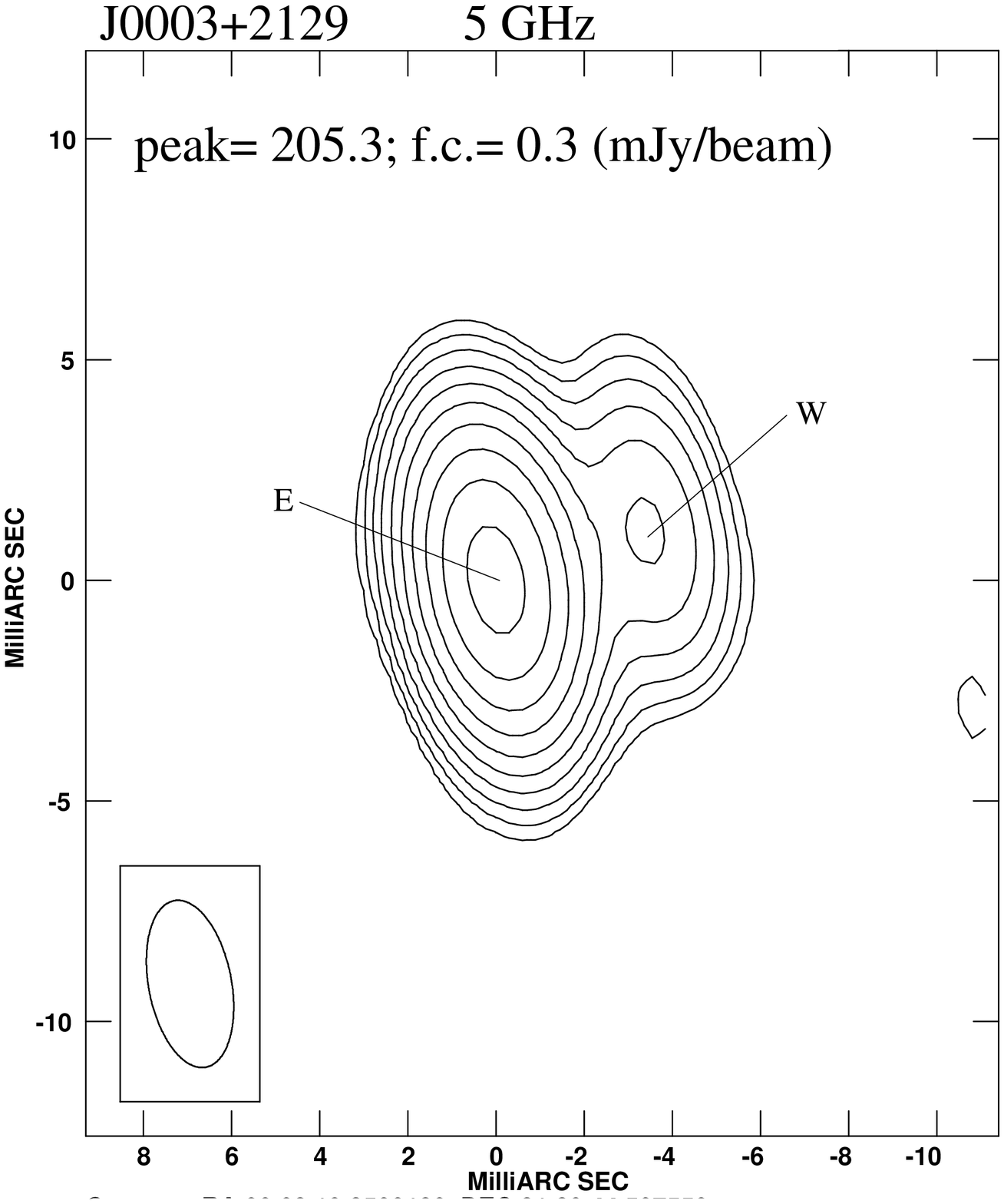}
\includegraphics{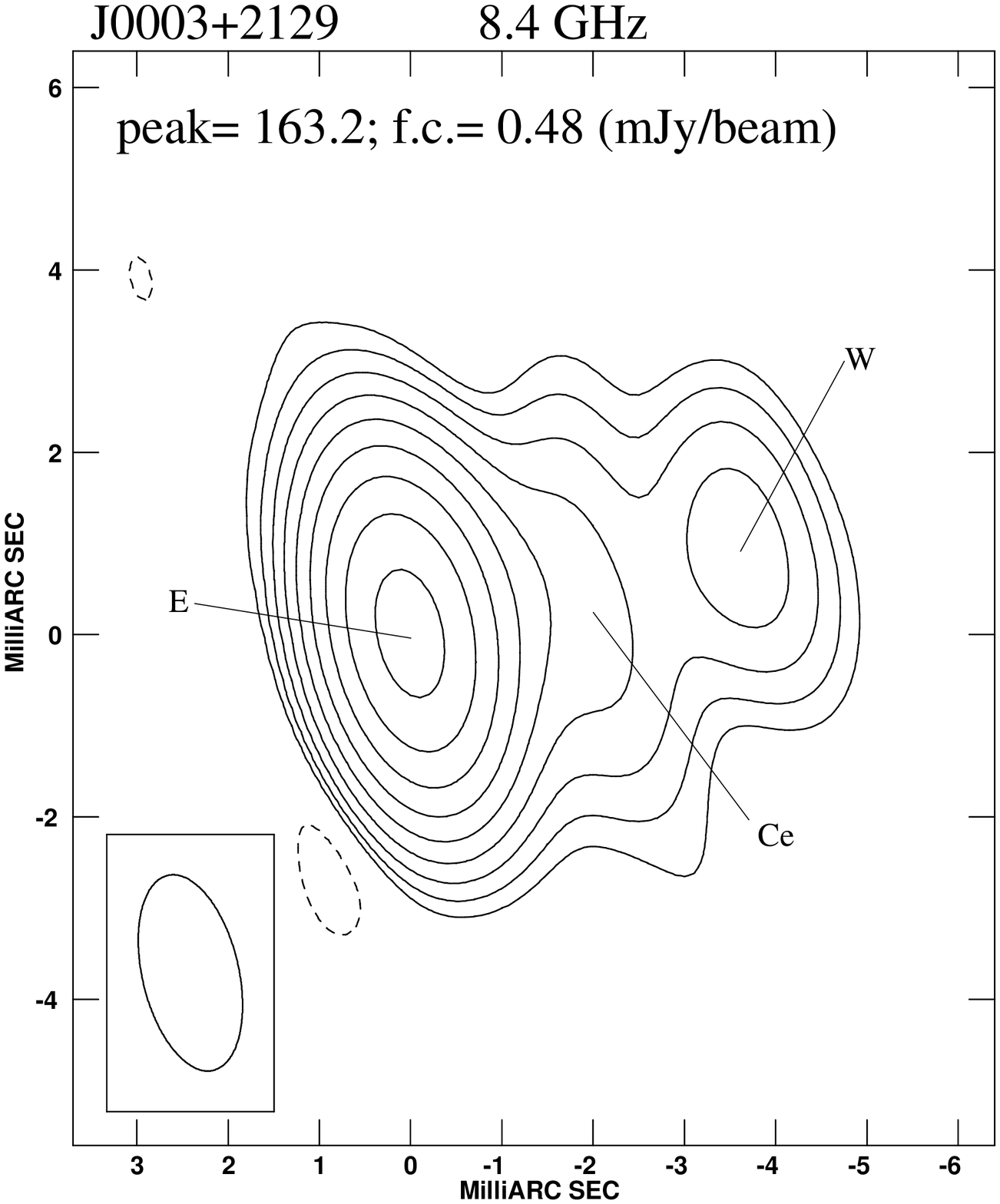}
\includegraphics{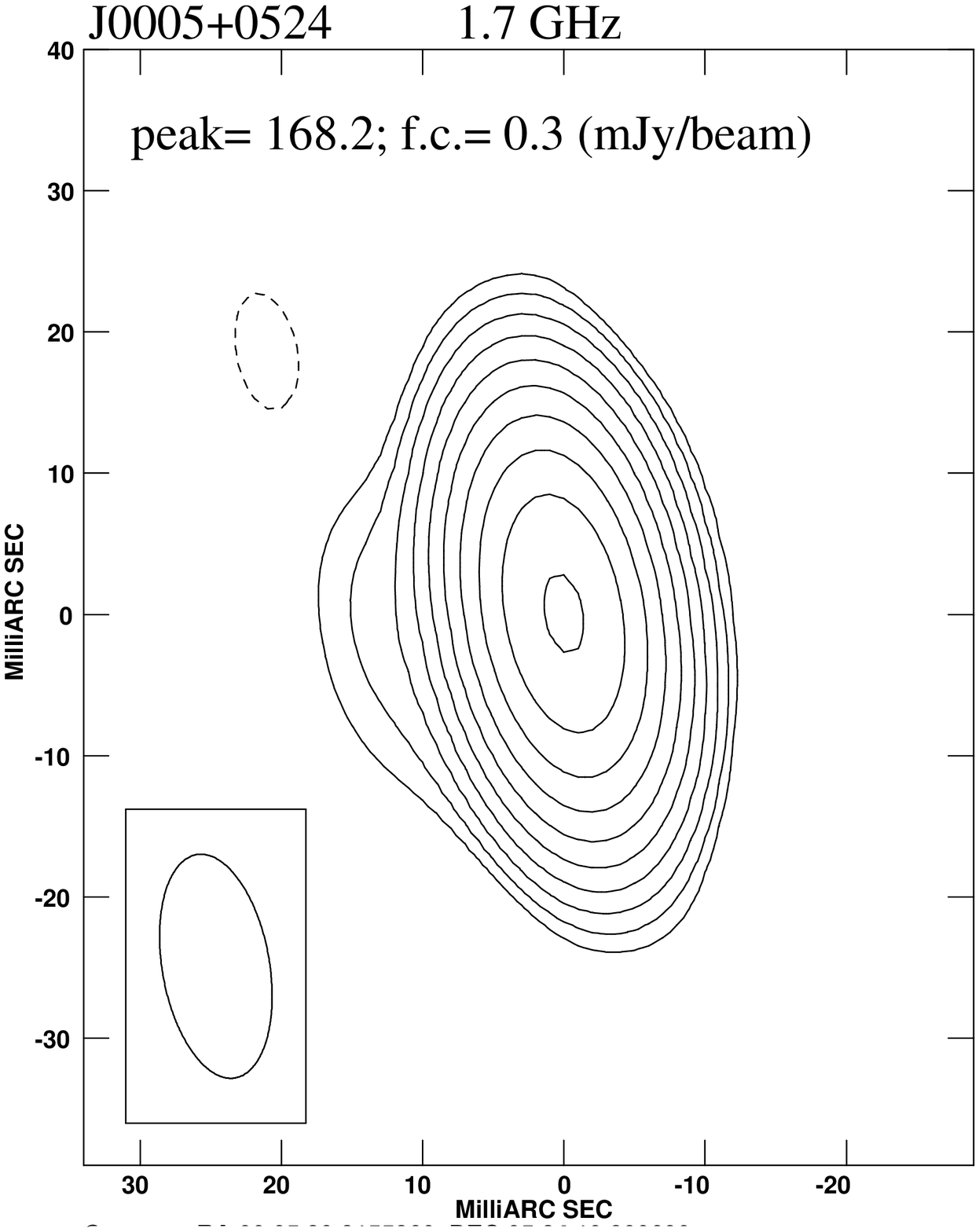}
\includegraphics{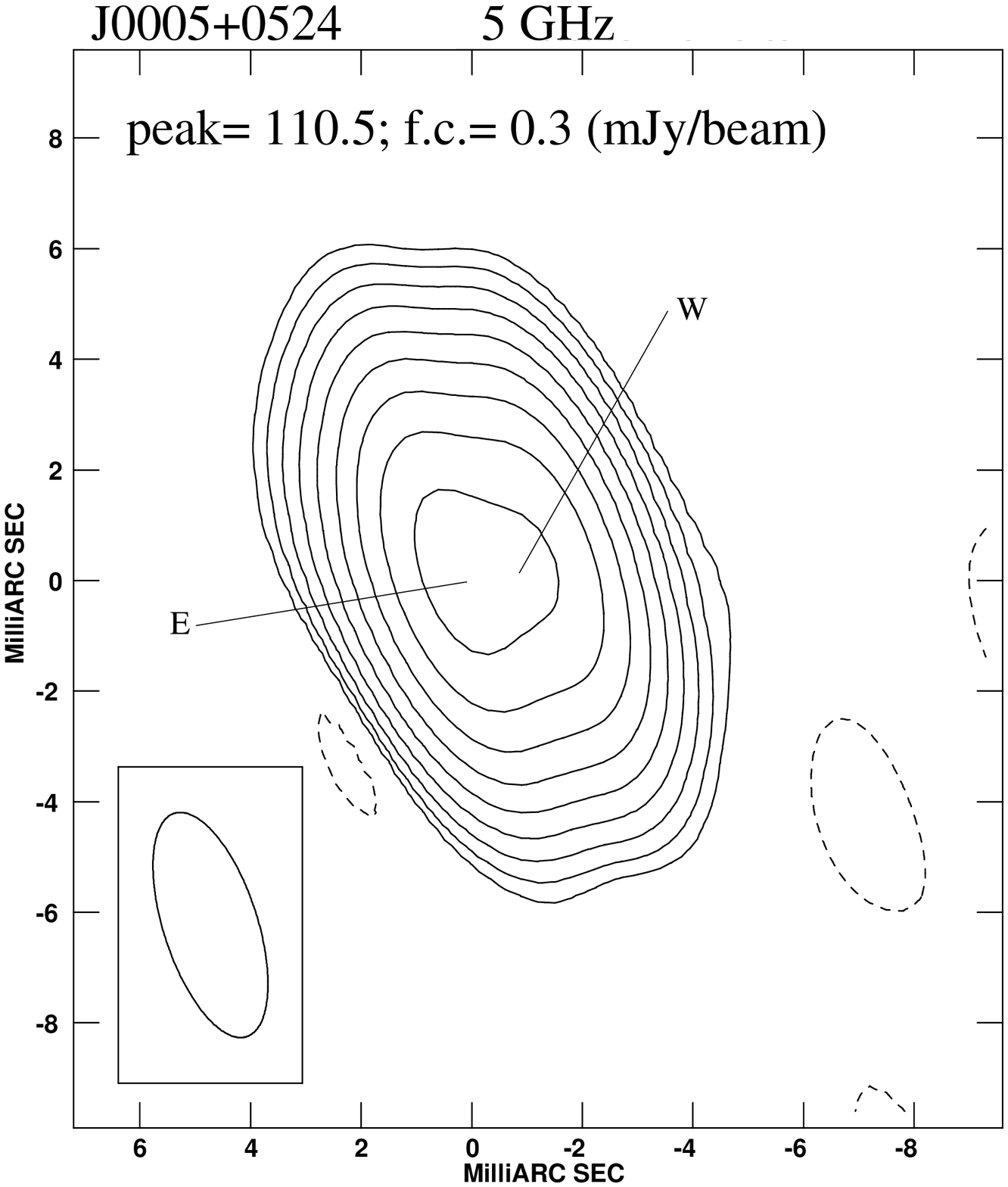}
\includegraphics{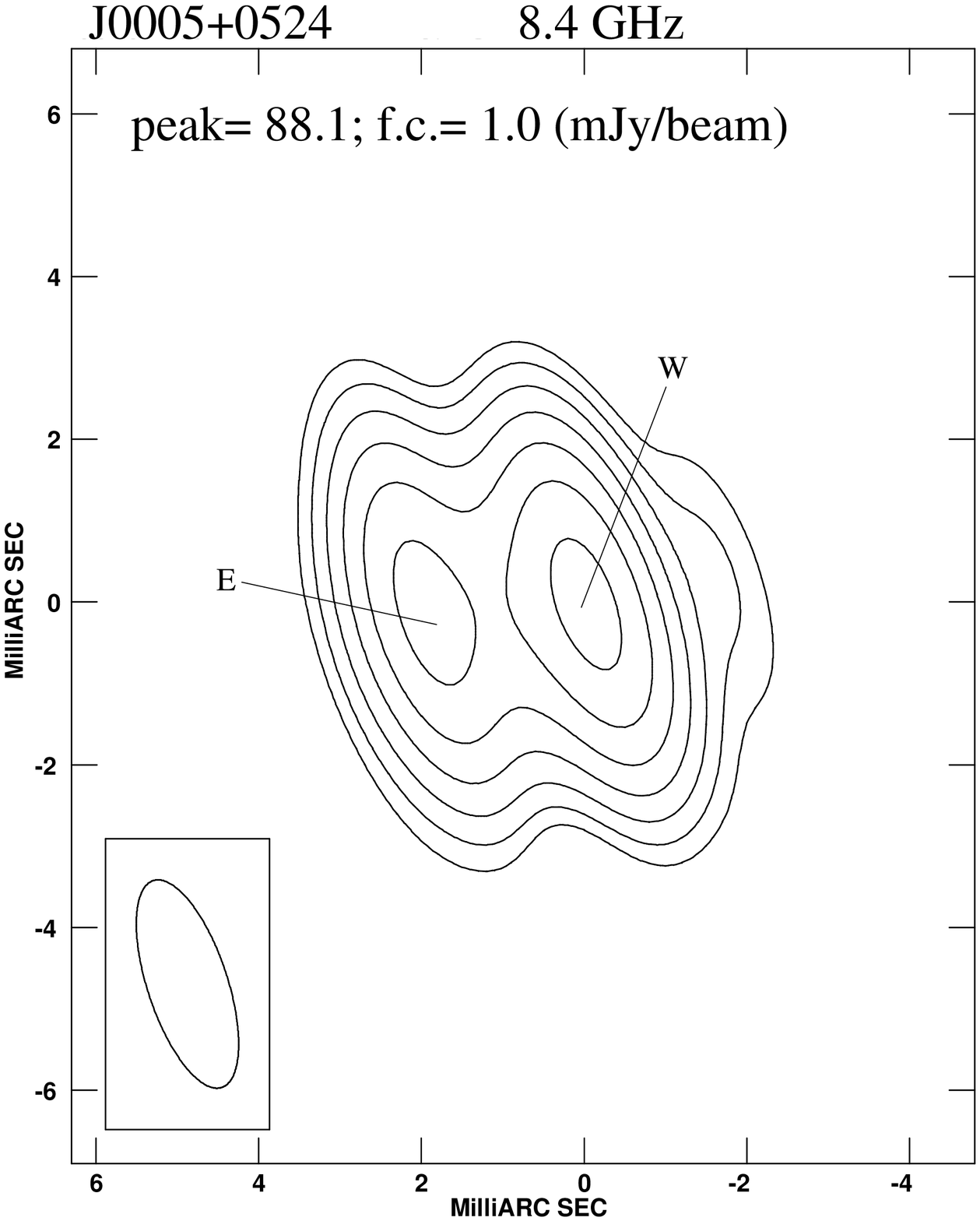}
\vspace{14cm}
\caption{VLBA images of the sources J0003+2129 ({\it top}) and J0005+0524
({\it bottom}) in L, C and X bands. VLBA observations of these two sources 
were carried out on July 22 2006.
 For each image, we provide the
  following information on the plot itself: {\bf a)} the source name
  and the observing frequency on the top left corner; {\bf b)} the peak
flux density in mJy/beam; {\bf c)} the first contour intensity ({\it
  f.c.} in mJy/beam), which is usually 3 times the off-source rms
noise level measured on the image; contour levels increase by a factor
2; {\bf d)} the restoring beam, plotted on the bottom left corner.}
\label{fig_00}
\end{center}
\end{figure*}

\begin{figure*}
\begin{center}
\includegraphics{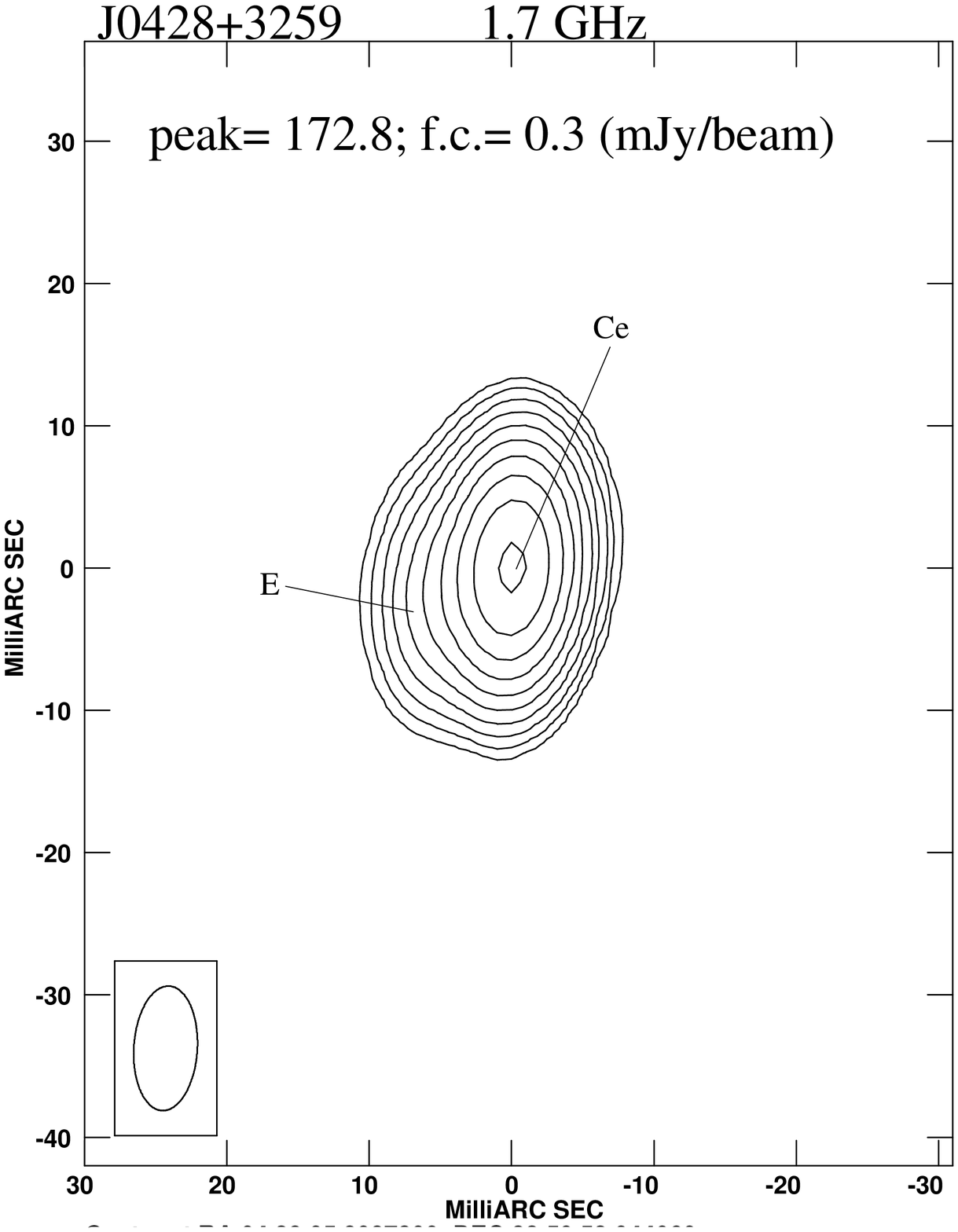}
\includegraphics{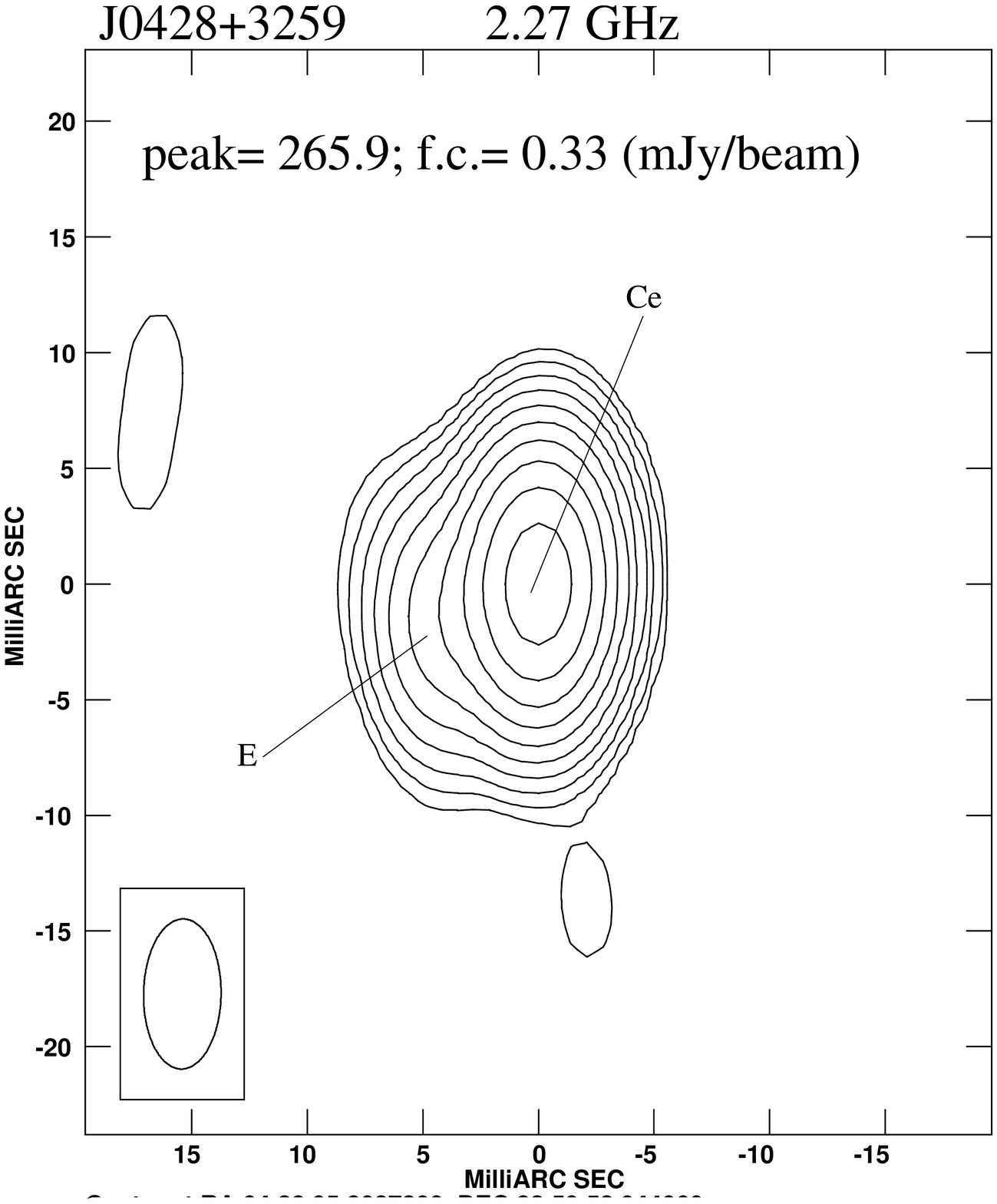}
\includegraphics{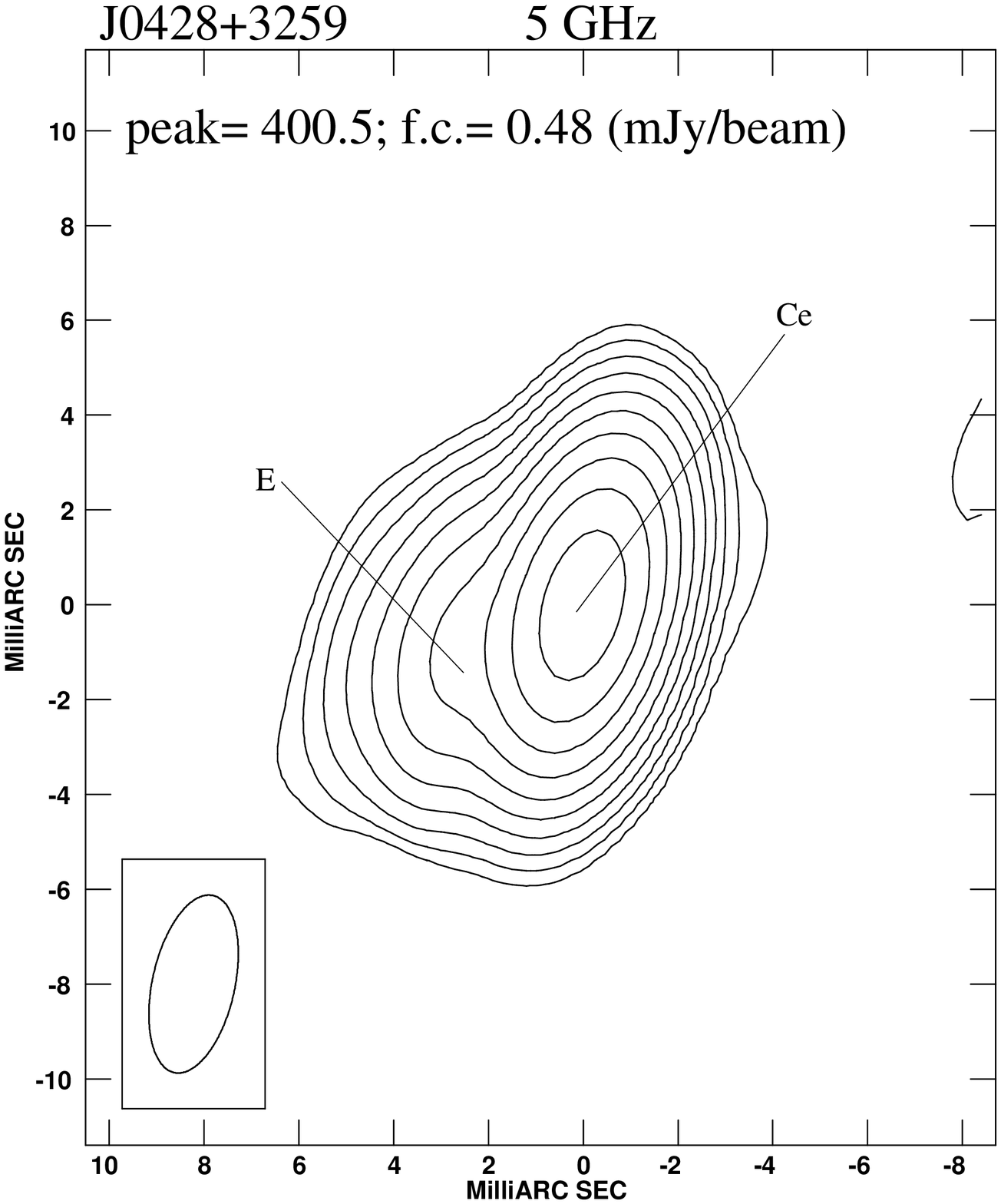}
\includegraphics{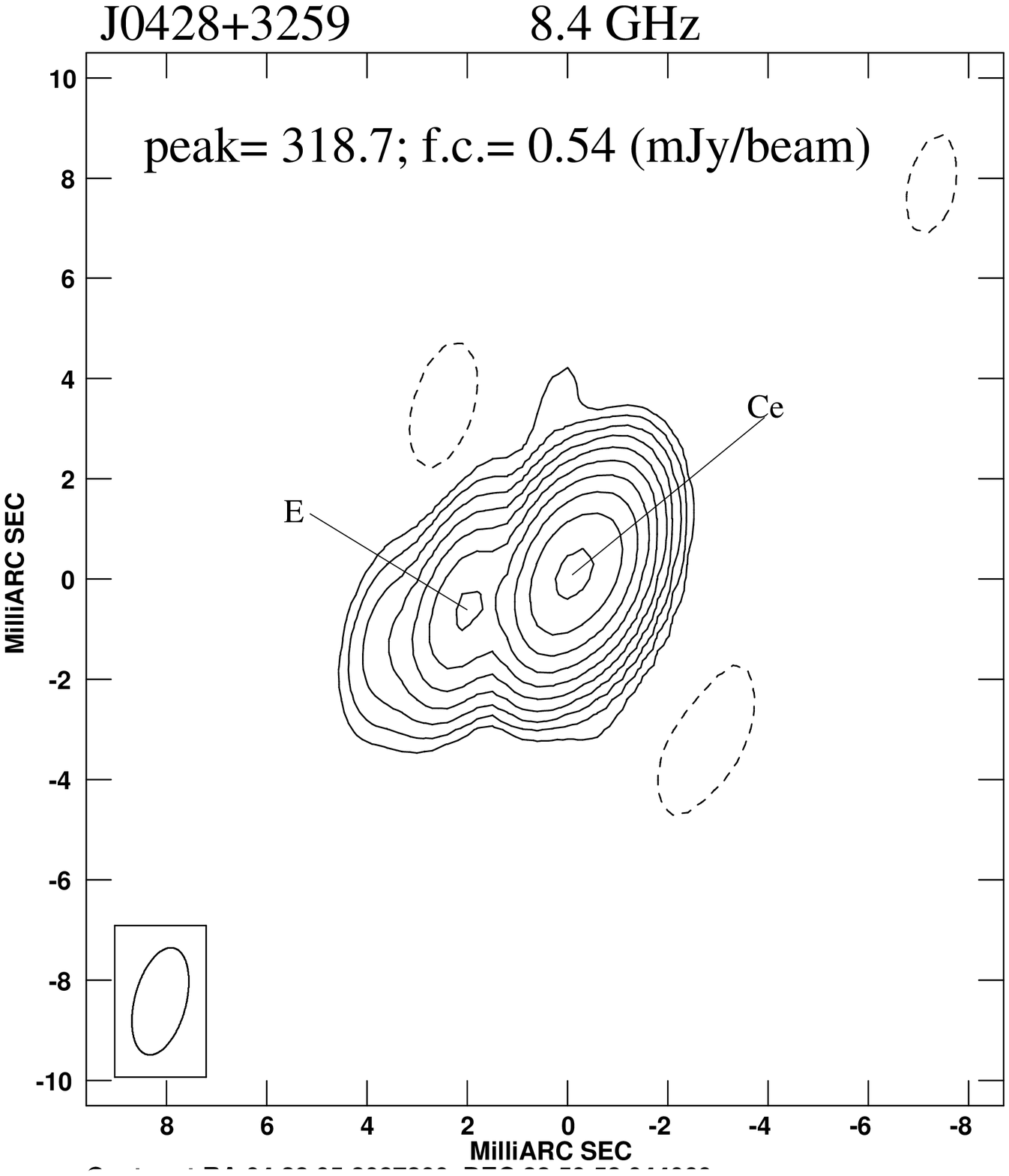}
\includegraphics{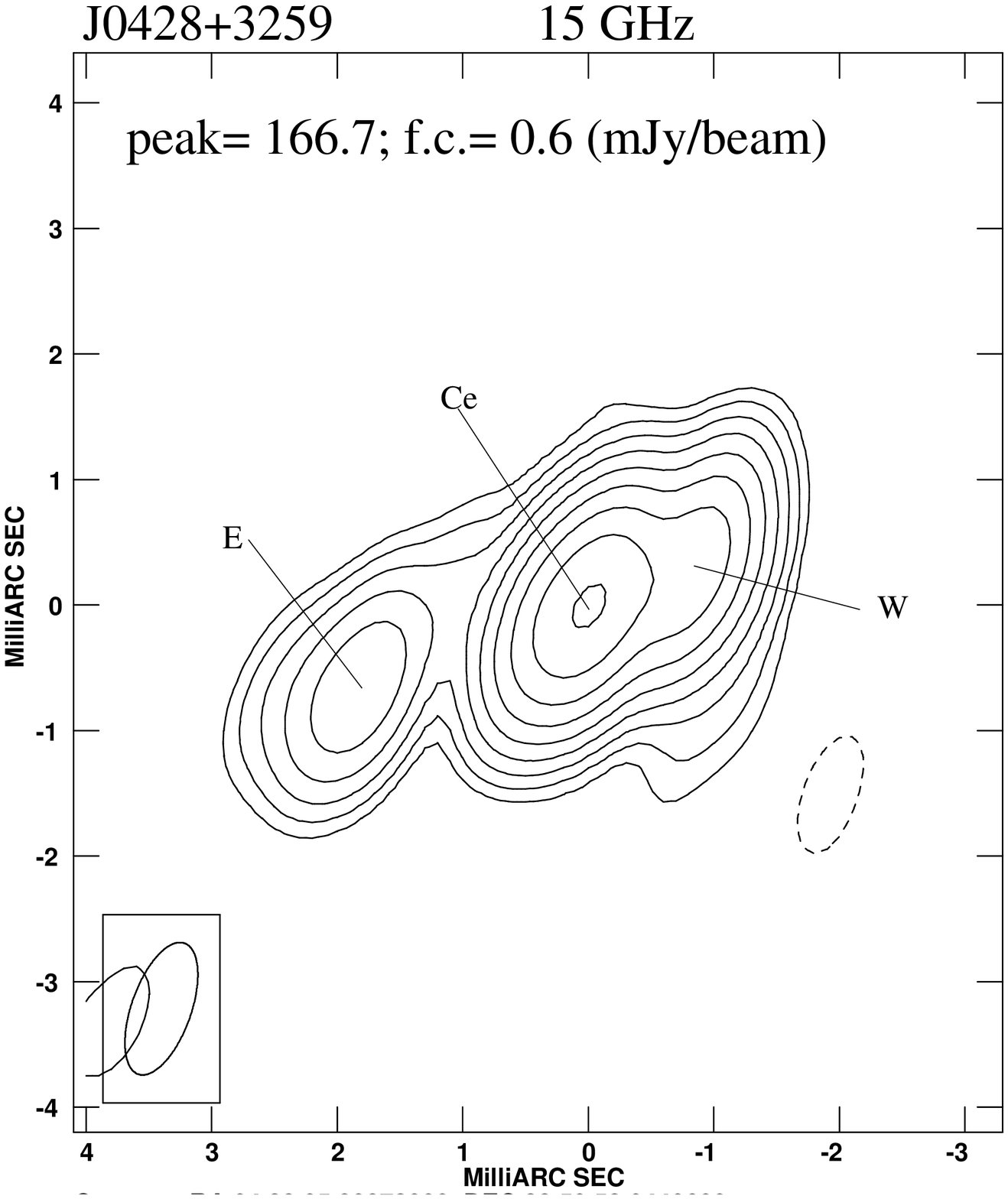}
\vspace{14cm}
\caption{VLBA images of the source J0428+3259 in L, S, C, X, and U
  bands. This source was observed with the VLBA on August 17 2006.
For each image, we provide the
  following information on the plot itself: {\bf a)} the source name
  and the observing frequency on the top left corner; {\bf b)} the peak
flux density in mJy/beam; {\bf c)} the first contour intensity ({\it
  f.c.} in mJy/beam), which is usually 3 times the off-source rms
noise level measured on the image; contour levels increase by a factor
2; {\bf d)} the restoring beam, plotted on the bottom left corner.}
\label{fig_0428}
\end{center}
\end{figure*}

\begin{figure*}
\begin{center}
\includegraphics{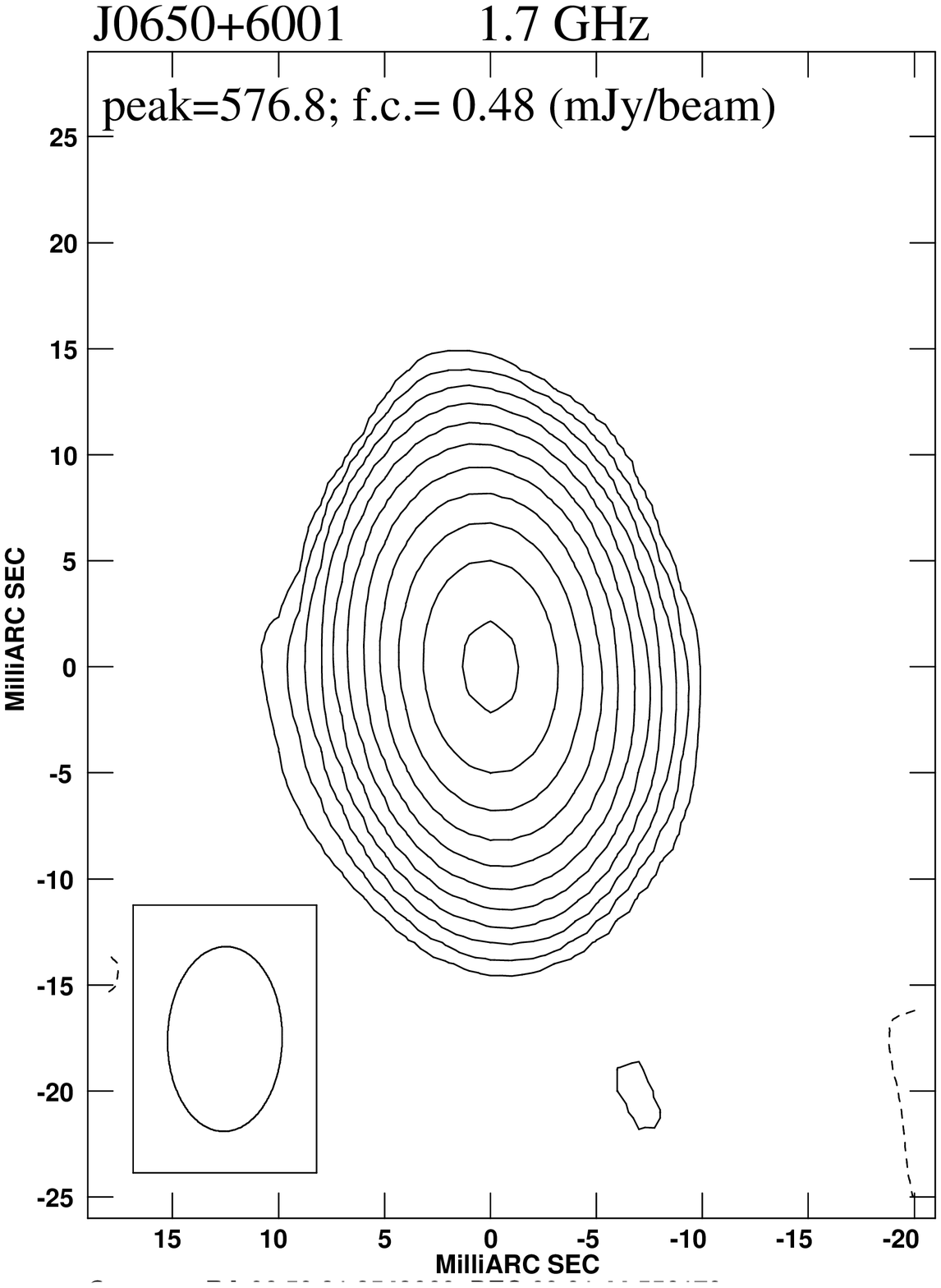}
\includegraphics{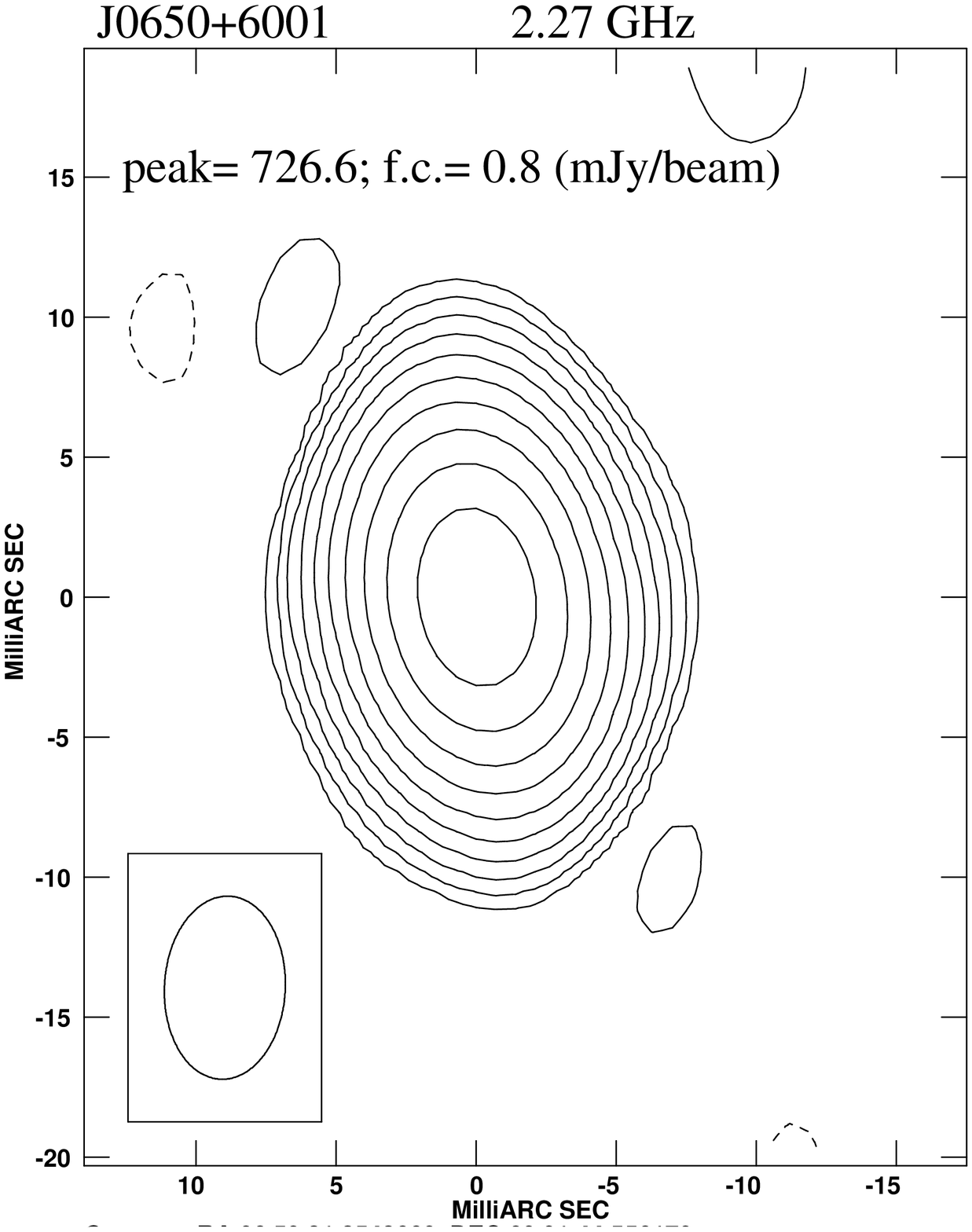}
\includegraphics{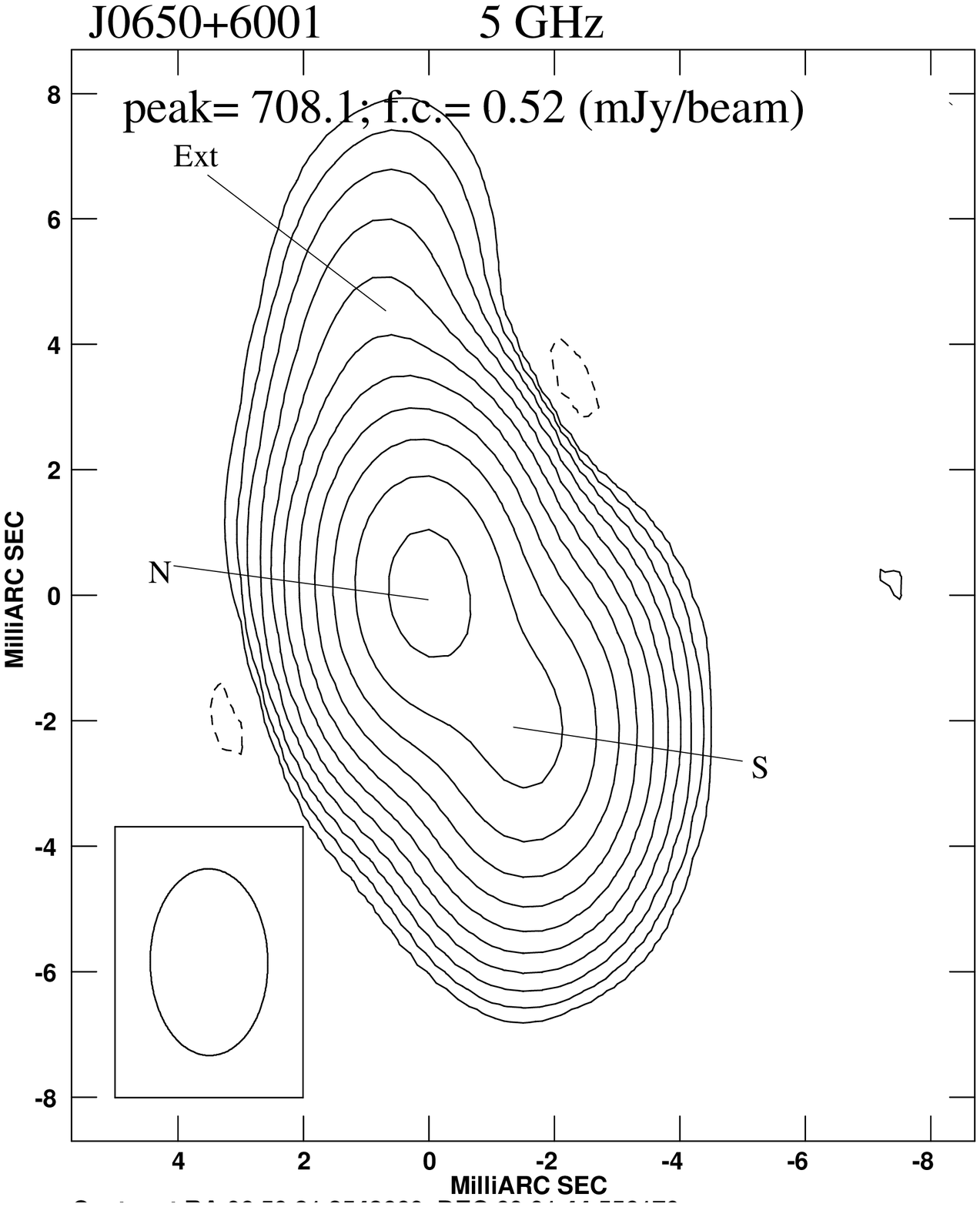}
\includegraphics{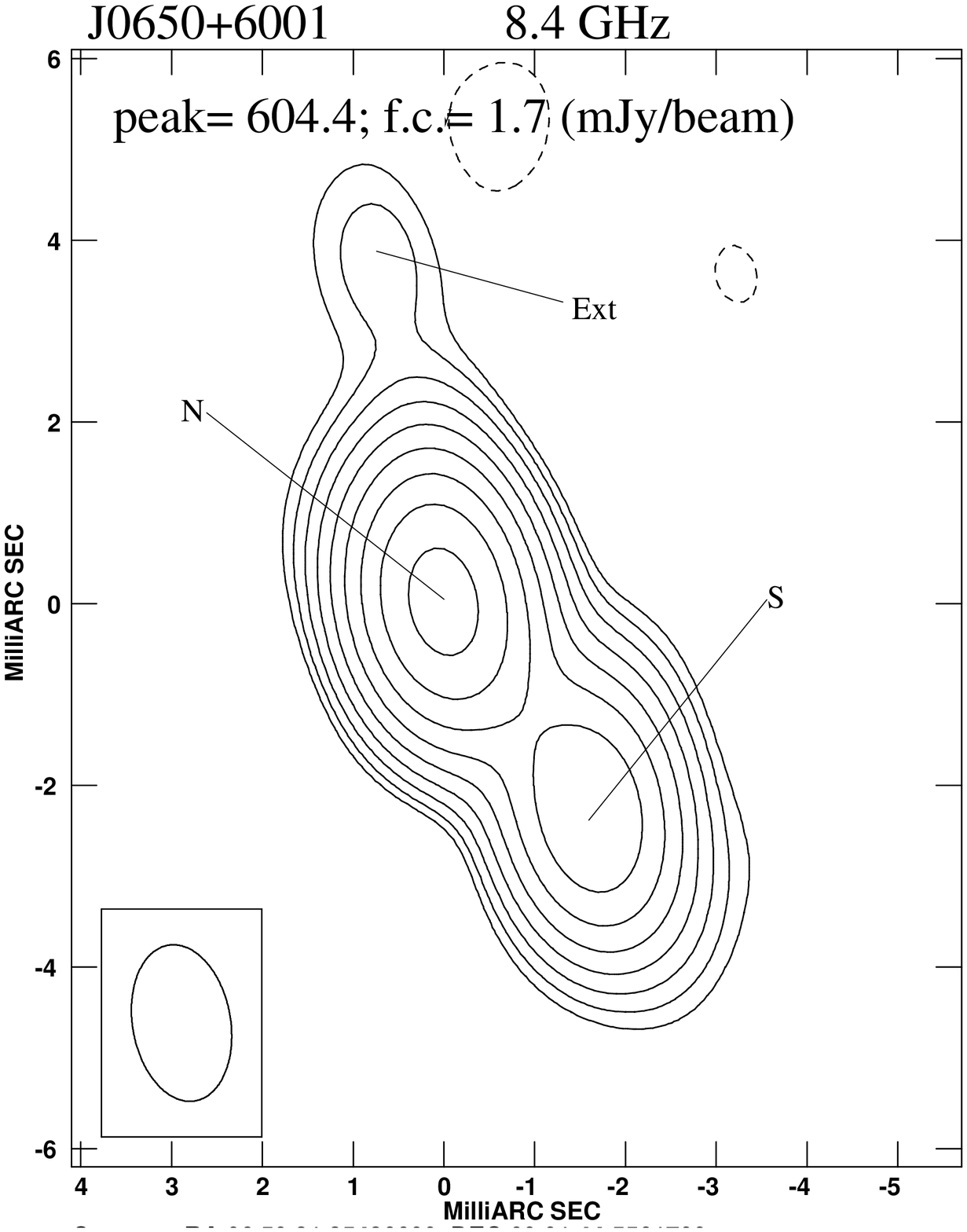}
\includegraphics{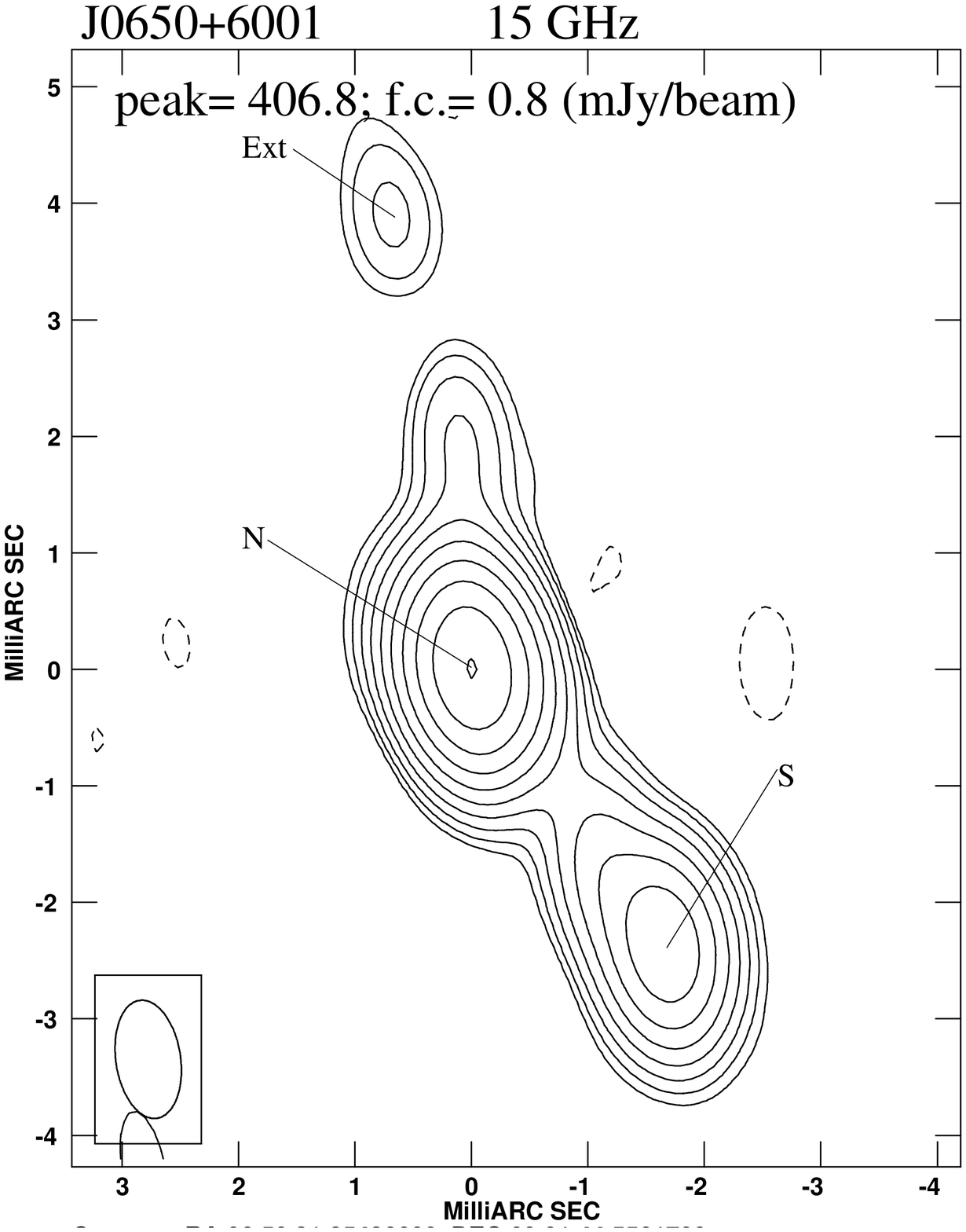}
\vspace{14cm}
\caption{VLBA images of the source J0650+6001 in L, S, C, X and U
  bands. VLBA observations were carried out on July 28 2006.
For each image, we provide the
  following information on the plot itself: {\bf a)} the source name
  and the observing frequency on the top left corner; {\bf b)} the peak
flux density in mJy/beam; {\bf c)} the first contour intensity ({\it
  f.c.} in mJy/beam), which is usually 3 times the off-source rms
noise level measured on the image; contour levels increase by a factor
2; {\bf d)} the restoring beam, plotted on the bottom left corner.}
\label{fig_0650}
\end{center}
\end{figure*}

The target sources were observed with the VLBA 
at 1.7 (L band), 5.0 (C band), and 15.3 (U band) 
in full polarization mode
with a recording band width of 16 MHz at 128 Mbps, while at 2.27 (S band)
and
8.4 GHz (X band) the dichroic receiver allowed a single polarization.
Observations were carried out between July and August 2006 in five
different runs, for a total time of 36 hours.
The correlation was performed 
at the VLBA correlator in Socorro and the data reduction was carried
out with the NRAO AIPS package. After the application of system
temperatures and antenna gains, the amplitudes were checked using data
on either 4C39.25 (J0927+3902) or DA\,193 (J0555+3948), both sources
having the flux density monitored at the VLA at 4.9 and 8.4 
GHz, allowing the verification of the amplitude calibration. The error
in the absolute flux density scale is generally within 3\%-10\%, being
highest in value at the highest frequency. \\
In C and U bands, the instrumental polarization was removed by using
the AIPS task PCAL; the absolute orientation of the electric vector of
DA\,193 and 4C\,39.25 was compared with the VLA/VLBA polarization
calibration database to derive the proper corrections. The values derived
from the two sources were in good agreement ($<$ 5$^{\circ}$).\\ 
For the sources J0003+2129 and J0005+0524, the system temperature
measured in the S band showed large variations and erratic values on
short timescales, probably due to interference, and no reliable images
could be obtained. 
At 15 GHz, the quality of data for these two sources, due to bad
atmospheric transmittance, is low and significantly worse than the
observations of Orienti et al. (\cite{mo06}).\\
The final images were obtained after a number of self-calibration
iterations. Amplitude self-calibration was applied only once at the
end of the process, using particular care; the solution interval (30 min)
was chosen to be longer than the scan-length to remove
residual systematic errors and fine tune the flux-density scale, but
not to force the individual data points to follow the model.\\
Full resolution images of the 5 HFP sources at the various
frequencies are presented in Figs. \ref{fig_00}, \ref{fig_0428},
\ref{fig_0650}, and \ref{fig_1511}. For each figure we provide the
following information: the source name and the observing frequency on
the top left corner; the peak flux density in mJy/beam; the first
contour intensity ({\it f.c.} in mJy/beam), which is usually 3 times
the off-source rms noise level measured on the image; contour levels
increase of a factor 2; the restoring beam, plotted on the bottom left
corner.\\
The final rms noises (1$\sigma$) are generally between 0.1 and 0.5
mJy/beam. 
The dynamic range, defined as the ratio of
the peak brightness to 1$\sigma$, is usually between 1000 and 3000.\\

\begin{table*}
\begin{center}
\caption{Observational parameters of the HFP source components.}
\begin{tabular}{c|c|c|c|c|c|c|c|c|c|c|c|c|c|c}
\hline
Source&Comp.&Opt&z&$S_{\rm 1.4}$&$S_{\rm 2.27}$&$S_{\rm 5.0}$&$S_{\rm
  8.4}$&$S_{\rm 15}$&$S_{\rm 22}$&$\nu_{p}$&S$_{p}$&$\theta_{\rm maj}$& $\theta_{\rm
  min}$&PA\\
 & & & &mJy&mJy&mJy&mJy&mJy&mJy&GHz&mJy&mas&mas&$^{\circ}$\\
(1)&(2)&(3)&(4)&(5)&(6)&(7)&(8)&(9)&(10)&(11)&(12)&(13)&(14)&(15)\\
\hline
&&&&&&&&&&&&&&\\
J0003+2129&E&G&0.45&101& &214&168&126$^{a}$& &5.0$\pm0.6$&220&0.41&0.12&140\\
 &Ce& & & & & &23& & & & &2.35&1.15&68\\
 &W& & & & &15&7&5$^{a}$& & & &0.78& - &170\\
J0005+0524&E&Q&1.887&173& &110&60&23$^{a}$& & & &0.81&0.80&97\\
 &W& & & & &79&103&82$^{a}$& & & &0.66&  -& -\\
J0428+3259&Ce&G&0.479&183&278&453&478&250&168$^{a}$&5.9$\pm0.1$&500&0.62&0.17&107\\
 &E& & &12&21&52&40&27&12$^{a}$&6.2$\pm$0.9&60&0.70&0.21&118\\
 &W& & & & & & &72&40$^{a}$& & &0.32&0.25&150\\
J0650+6001&N&Q&0.455&637&866&745&658&495&327$^{a}$&6.35$\pm$0.01&770&0.37&0.24&25\\
 &S& & & & &337&253&122&78$^{a}$&3.1$\pm$0.1&460&0.56&0.31&28\\
 &Ext& & & & &26&5&4& & & &2.3&0.4&20\\
J1511+0518&E&G&0.084& &27&127&224&242&150$^{a}$&11.23$\pm$0.46&250&0.43&0.16&80\\
 &Ce& & & & & &20&32& & & &0.27&0.13&60\\
 &W& & &94&191&484&458&449&285$^{a}$&7.1$\pm$0.1&500&0.56&0.13&70\\
 &Ext& & &3& & & & & & & &18&12&170\\
&&&&&&&&&&&&&&\\
\hline
\end{tabular}
\note{Observational parameters of the HFP source components. 
Column 1: source name (J2000); Column 2: component; Column 3: optical
identification: G=galaxy, Q=quasar; Column 4: redshift; 
Columns 5, 6, 7, 8, 9 and 10: VLBA flux density at
1.4, 2.3, 4.8, 8.4, 15.3 and 22.2 GHz; Columns 11, 12 and 13: the deconvolved
major and minor axis and position angle of the major axis. 
{\it a}: Flux density from Orienti et al. (\cite{mo06})}
\label{parameter}
\end{center}
\end{table*}

\section{Results}

Measurements of the source parameters with small
uncertainties allow us to reliably derive the physical 
conditions in the target sources supposed to be in the earliest phase of
the radio source evolution.\\
Source parameters, such as flux density and deconvolved angular sizes
were measured by means of the task JMFIT, which performs a
Gaussian fit on the image plane. The solutions were generally
rather good, providing residuals consistent with the off-source noise.
In case of extended components (e.g. J1511+0518 in L band),
the flux density was derived in the image plane 
by means of the tasks TVSTAT and IMSTAT. \\
Peak frequency and peak flux-density 
were determined by fitting the multi-frequency spectra
of each source sub-component with a pure analytical
function (see e.g. Dallacasa et al. \cite{dd00}). Where available, we
considered the 22-GHz data point from Orienti et
al. (\cite{mo06}) to constrain more accurately the fit at high
frequencies.
Errors in peak frequency and peak flux density were calculated 
by error propagation theory: the former are
reported in Table \ref{parameter}, while the latter account only for $\sim$
1-3\% of the peak flux density. 
In the case of the Eastern component of
J0005+0524, the fit could not constrain the peak of the radio spectrum 
with sufficient accuracy 
and we do not derive the magnetic field using
the parameters provided by the fit. \\
In Fig. \ref{spettro}, {\it crosses} indicate the VLBA spectra of 
sub-components; {\it plus signs} show the total VLBA flux density obtained
adding the spectra of each sub-component; {\it diamonds} represent the
VLA overall flux density of Orienti et al. (\cite{mo07}).  
For a few source components, the availability of 3 or less data points
did not allow us an accurate determination of the peak parameters, and
they were not considered any further.\\    
In Table \ref{parameter}, we report the observational parameters of each
source component.\\ 
In the case of J0650+6001, the flux density at 1.7 and 2.27 was
rescaled from Akujor et al. (\cite{akujor96}). Errors associated with
these extrapolations were conservatively to be $\sim$10\%.\\
No polarized emission was detected from the target sources at
the mJy level in C and U bands, which is consistent with VLA
measurements (Orienti \& Dallacasa \cite{mo08b}).\\

\subsection{Source morphology}

The high-resolution achieved by these multi-frequency VLBA
observations was adequate to resolve the HFP sources into a small number
of sub-structures, each with its own radio spectrum and a
rather accurate determination of source parameters (in particular the
angular size) allowing us to constrain directly the magnetic field in
a homogeneous component. \\
The observed sources were selected on the basis of their
CSO-like morphology as found by Orienti et al. (\cite{mo06}) by means
of observations in the optically-thin part of their radio
spectrum. These new observations, carried out at several frequencies 
below and above the spectral peak, provide more information on the
source sub-structures and their individual spectra.\\   
The morphologies derived from these new observations are consistent with
those found in previous observations, and a proper discussion of the
source structures can be found in Orienti et al. (\cite{mo06}).
We briefly summarize the information emerging from these new
observations.\\

\noindent The galaxy J0003+2129, found to be Double by
Orienti et al. (\cite{mo06}), displays a Triple structure in our
new observations. The higher resolution achieved at 8.4 GHz 
by the present data enabled us to 
identify a faint region located between the two main
components at a distance of 
$\sim$ 1 mas ($\sim$ 6 pc) from the brightest one. \\

\begin{figure*}
\begin{center}
\includegraphics{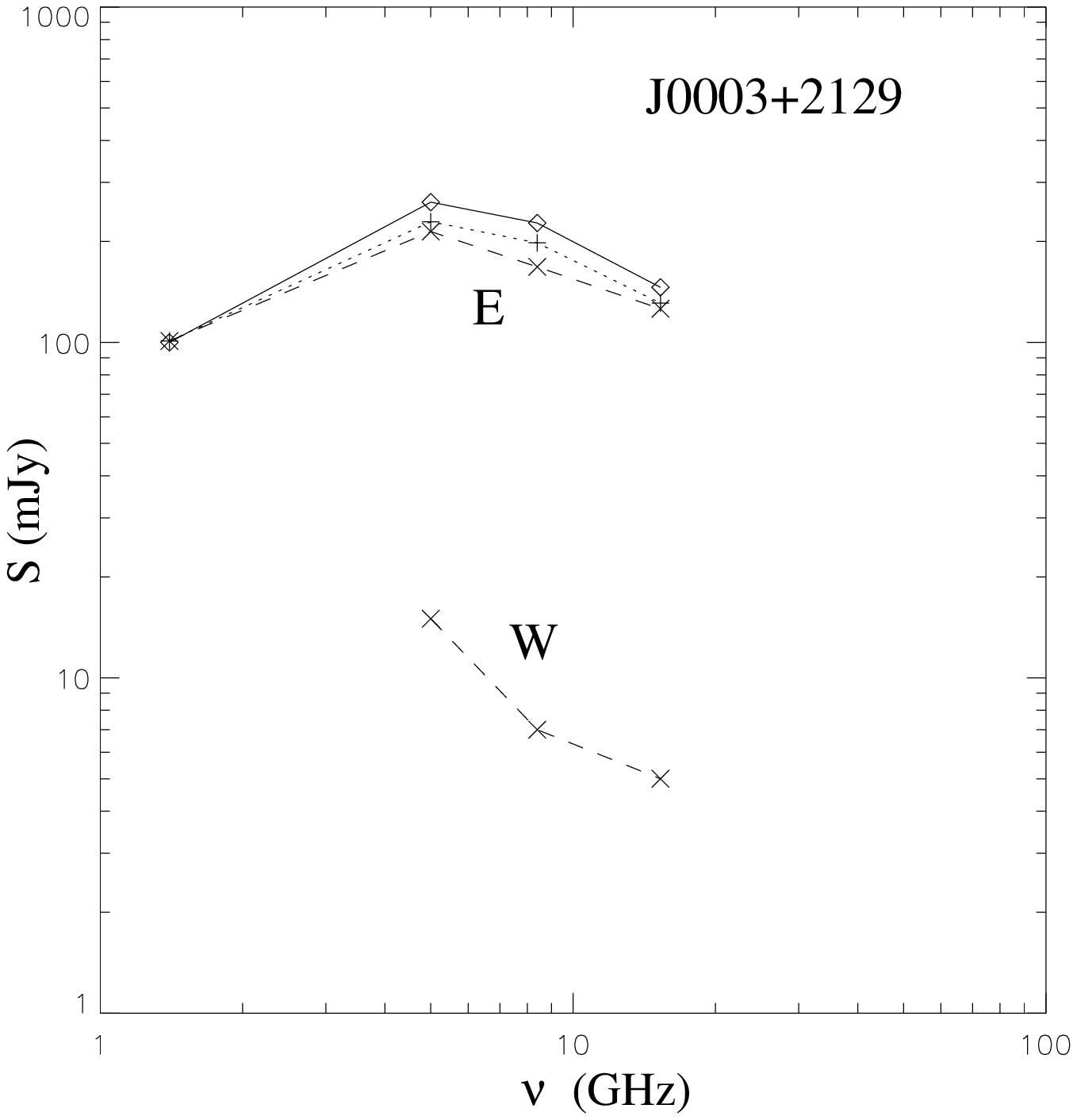}
\includegraphics{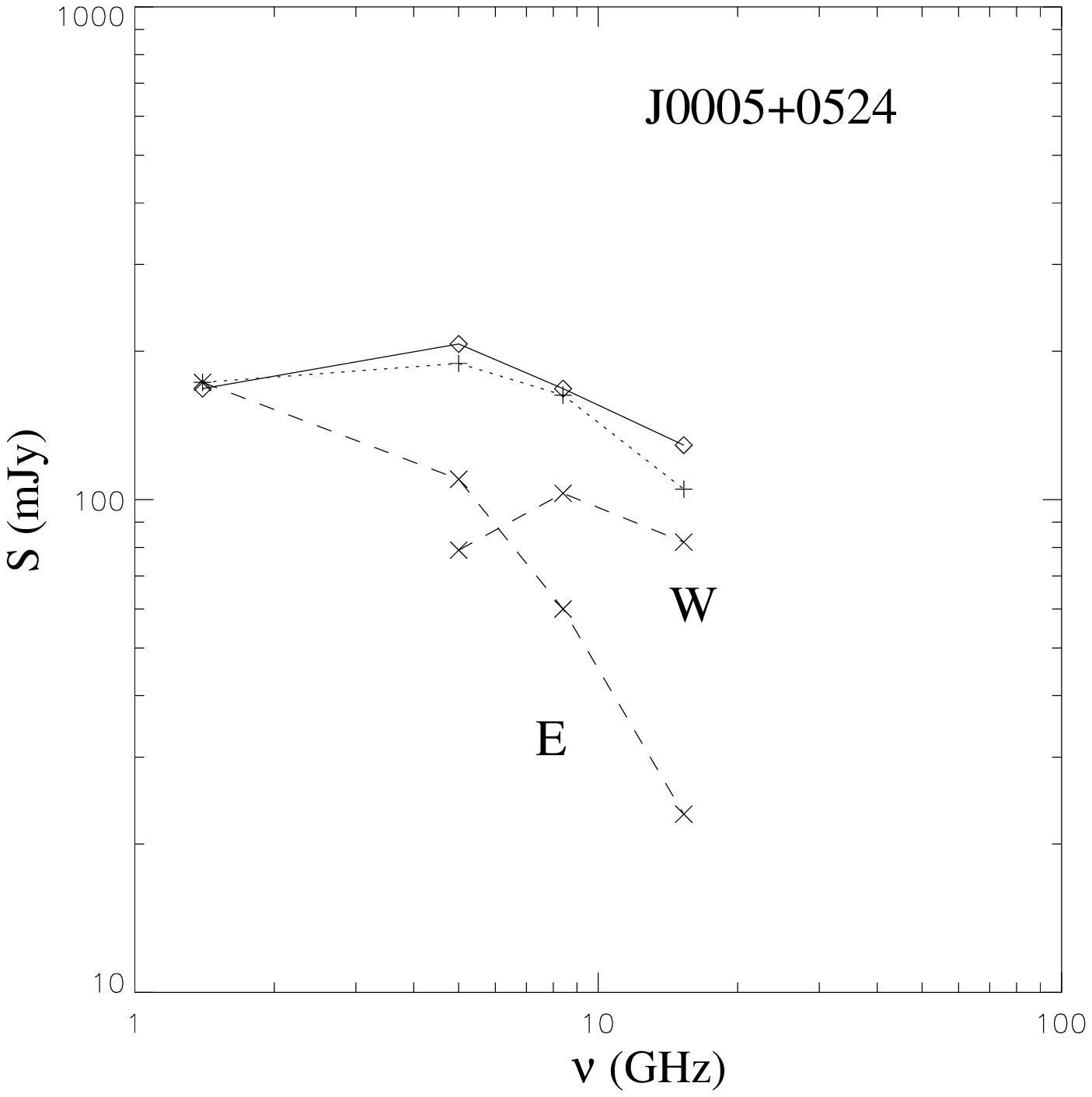}
\includegraphics{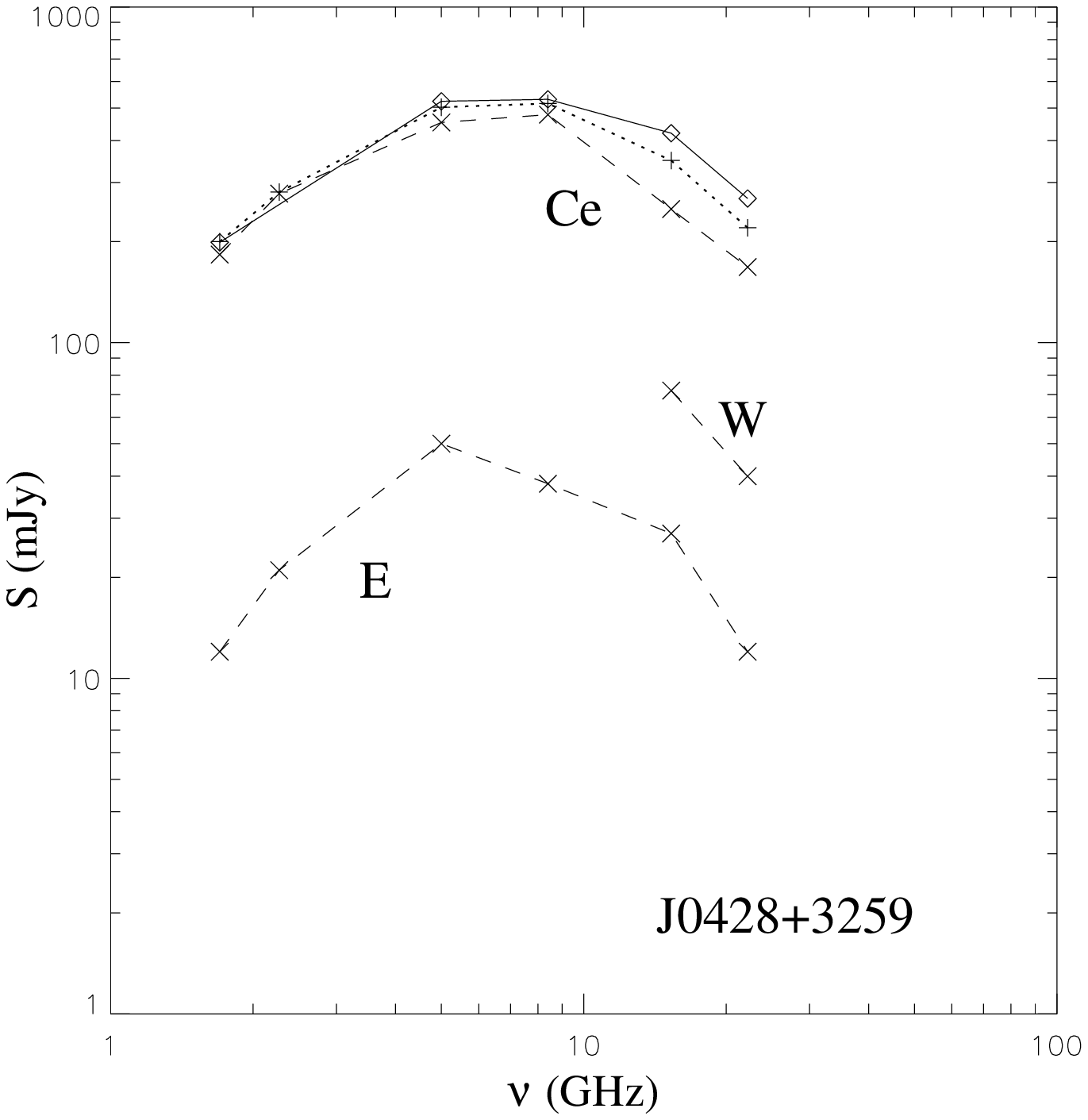}
\includegraphics{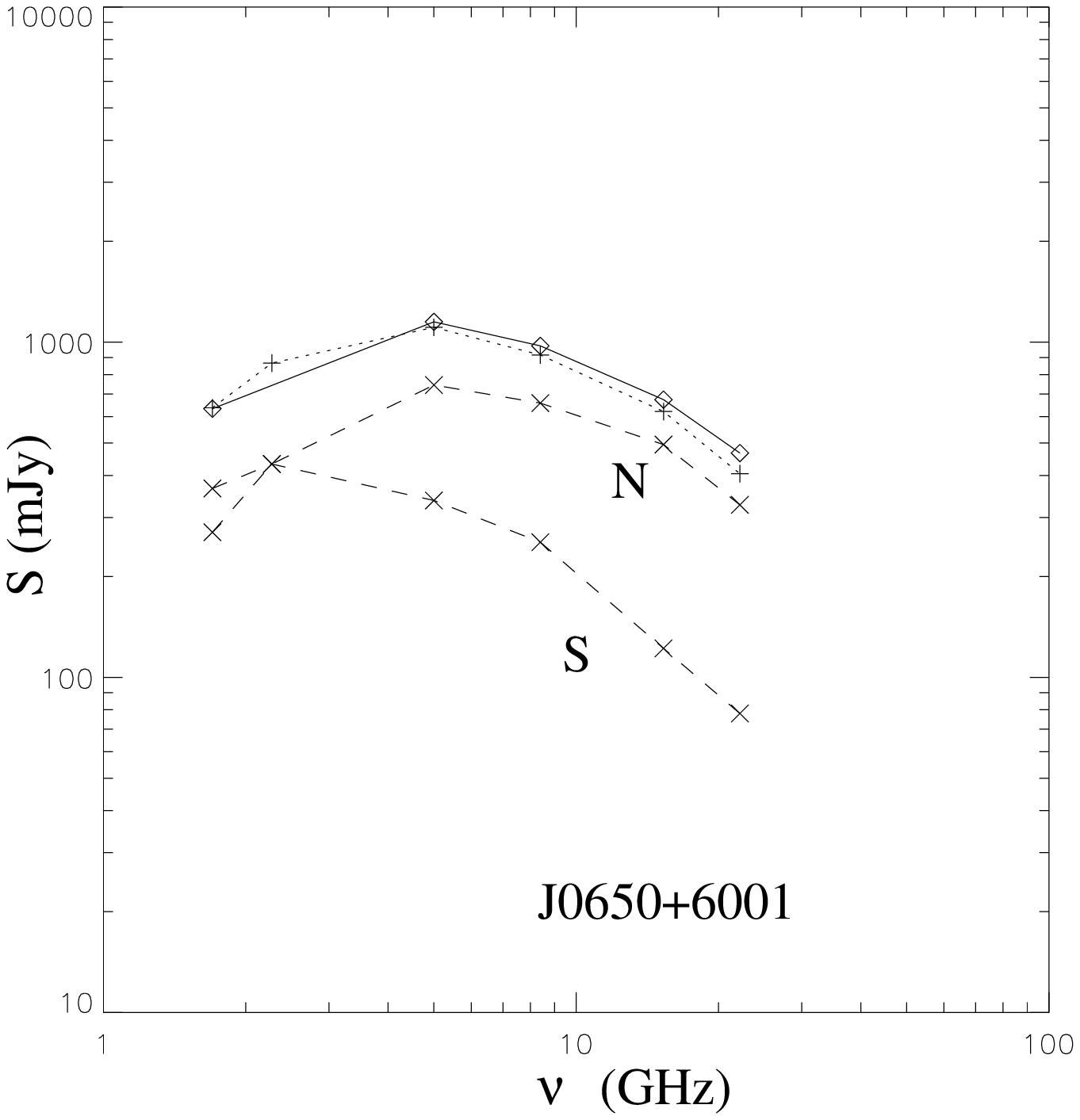}
\includegraphics{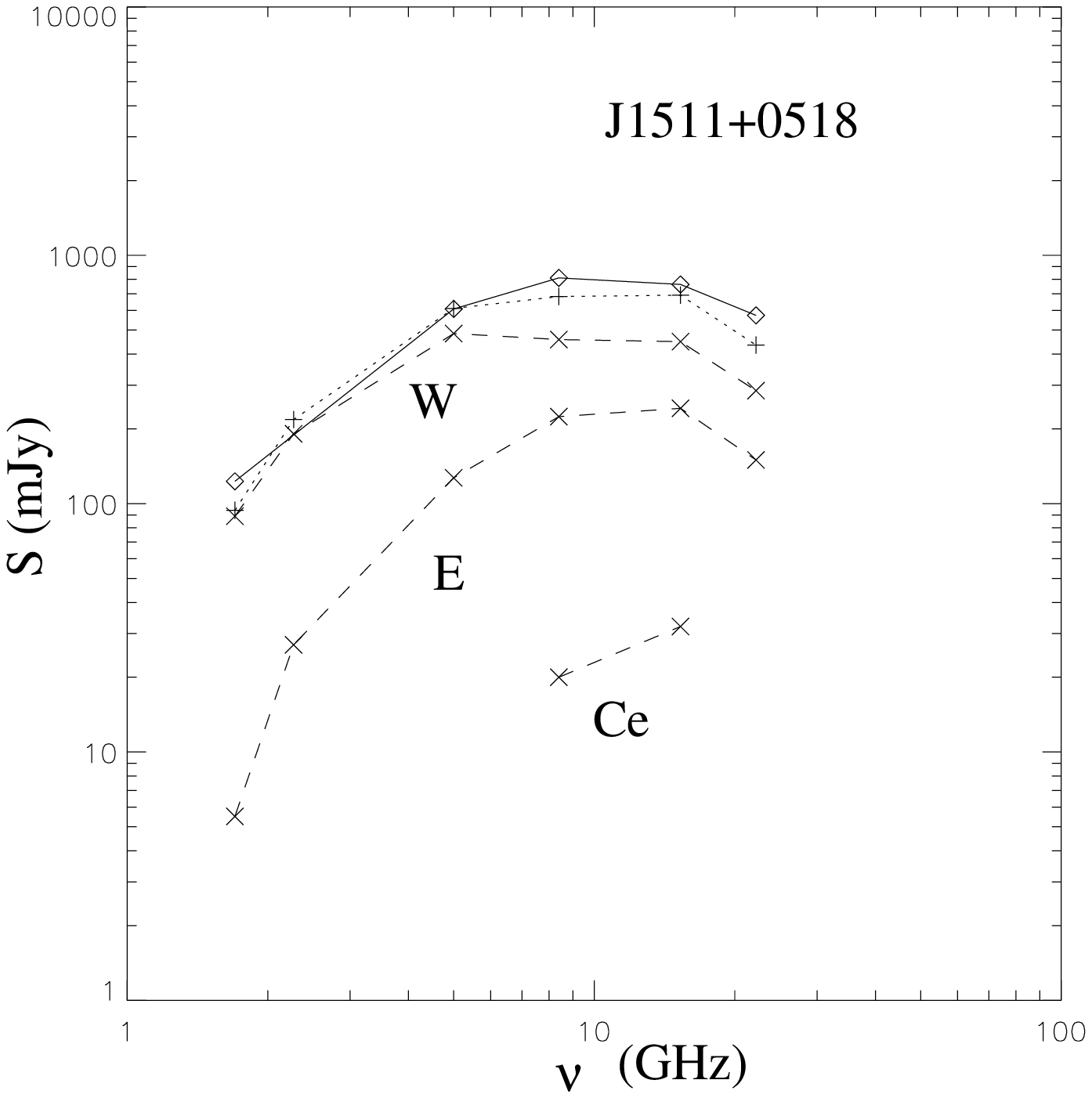}
\vspace{12.0cm}
\caption{Spectra of the 5 HFP sources and their sub-components. 
{\it Crosses} represent 
  VLBA spectra of sub-components; {\it plus signs} show the
  total VLBA flux density obtained adding the spectra of each 
  sub-component; {\it diamonds} represent the VLA overall flux density as
  from Orienti et al. \cite{mo07}.} 
\label{spettro}
\end{center}
\end{figure*}

\noindent The quasar J0005+0524 is characterized by a well
defined Double structure. However, the two components have different
characteristics: the spectrum of the Eastern
component peaks at lower frequency than the Western one
(Fig. \ref{spettro}), which is
still self-absorbed at 8.4 GHz and becomes optically-thin at $\sim$ 15 GHz
(Orienti et al. \cite{mo06}). At 1.4 GHz, there is a tentatively
detection (6$\sigma$) of an extended emission accounting for $\sim$ 3
mJy.\\

\noindent The galaxy J0428+3259 shows a well defined Triple
morphology, with a total linear size (LLS) of about 16 pc and a
position angle of 110$^{\circ}$. All the source components have steep
spectra and there is no indication of the source core.\\

\noindent The quasar J0650+6001 displays a Double structure, with an
extended emission located at $\sim$ 4 mas ($\sim$23 pc) 
and position angle of $\sim 10^{\circ}$ with respect to the
Northern component. 
Such an extended feature can be resolved at 5.0, 8.4, and 15 GHz only,
in agreement with other works with similar resolution (see
e.g. Stanghellini et al. \cite{cs99}; Akujor et al. \cite{akujor96}).\\

\noindent The galaxy J1511+0518 is characterized by a well defined
Triple structure. The availability of 8.4-GHz observations at a
higher spatial resolution than that achieved by Xiang et al. (\cite{xiang02})
allows us to identify the central component as the source core,
since it has an inverted spectral index $\alpha_{8.4}^{15} \sim
-0.8$. 
At low frequencies, we find an additional extended emission, accounting
for $\sim$ 3 mJy at 1.7 GHz, located at $\sim$ 30 mas ($\sim$ 50 pc)
and oriented at a different
position angle (p.a. $\sim 145^{\circ}$) with respect to the main source
structure  (p.a. $\sim 95^{\circ}$).\\

\begin{figure*}
\begin{center}
\includegraphics{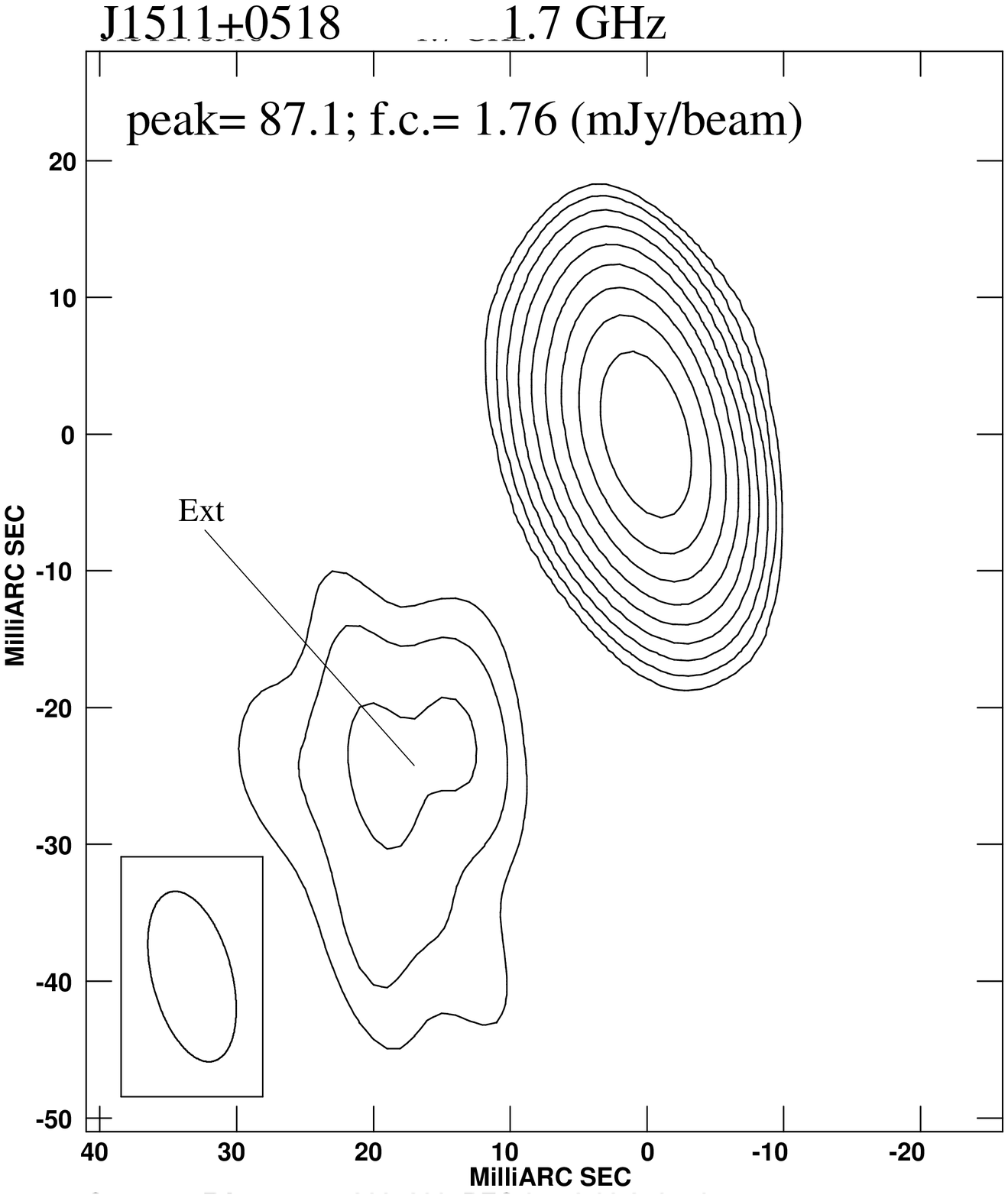}
\includegraphics{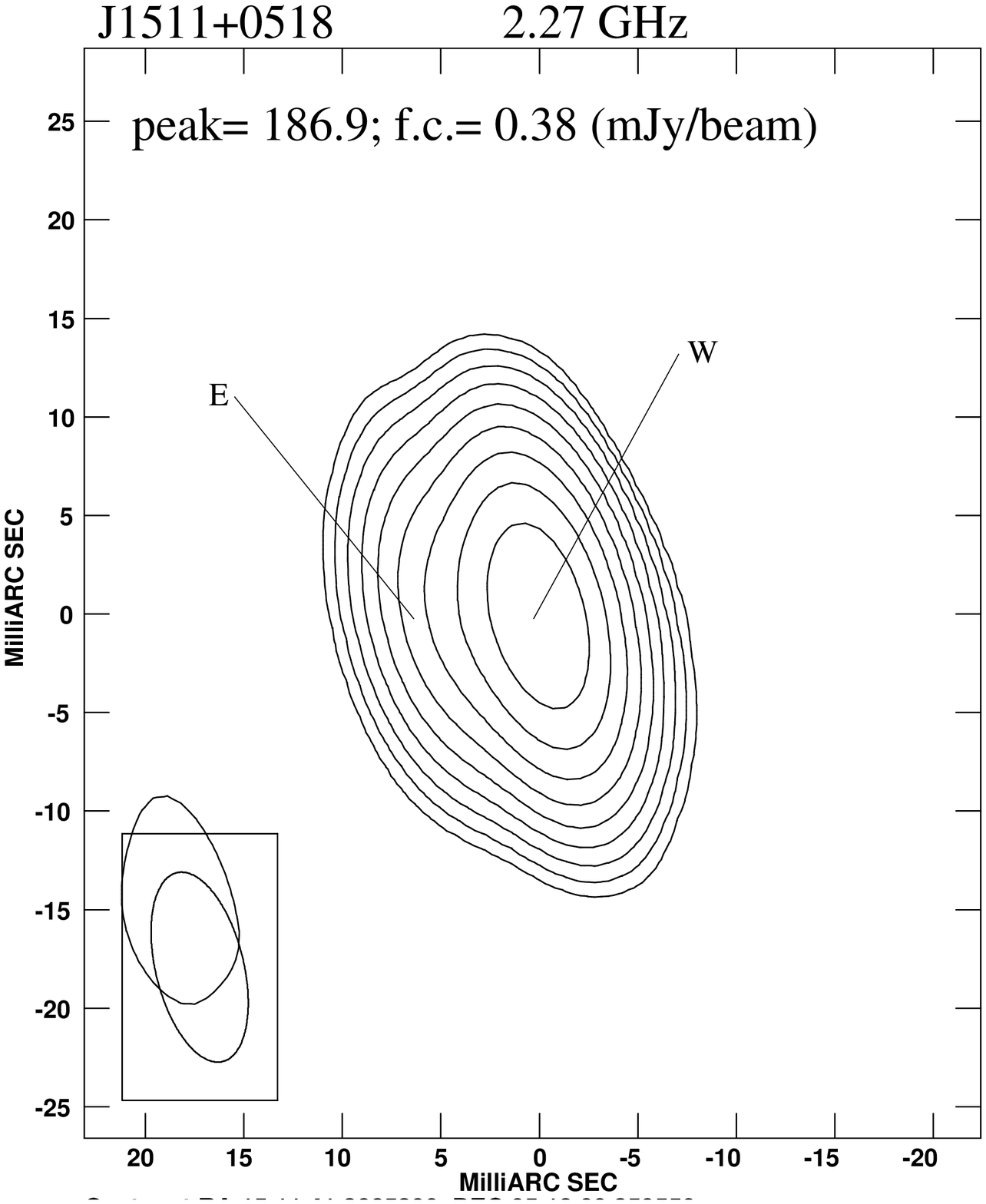}
\includegraphics{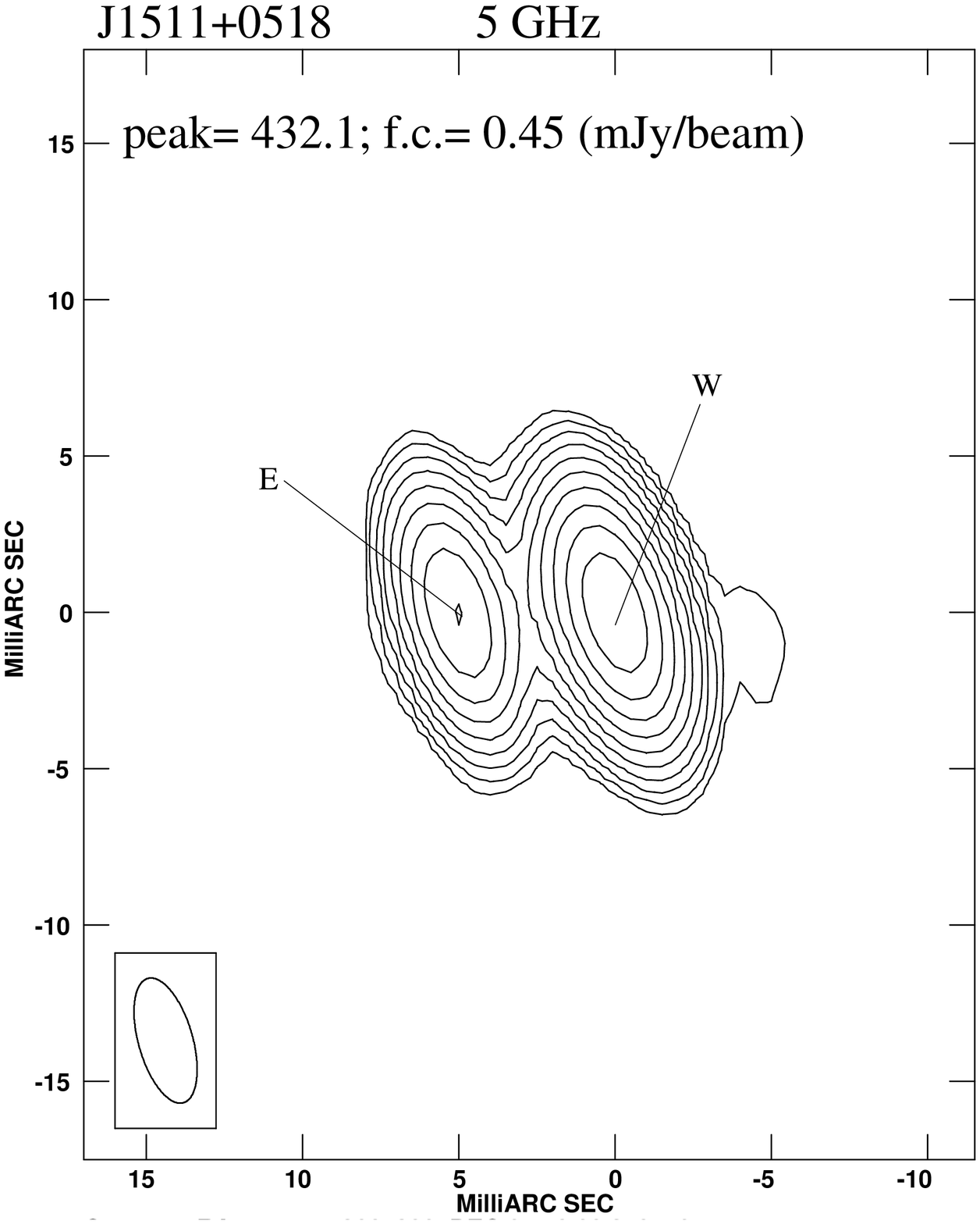}
\includegraphics{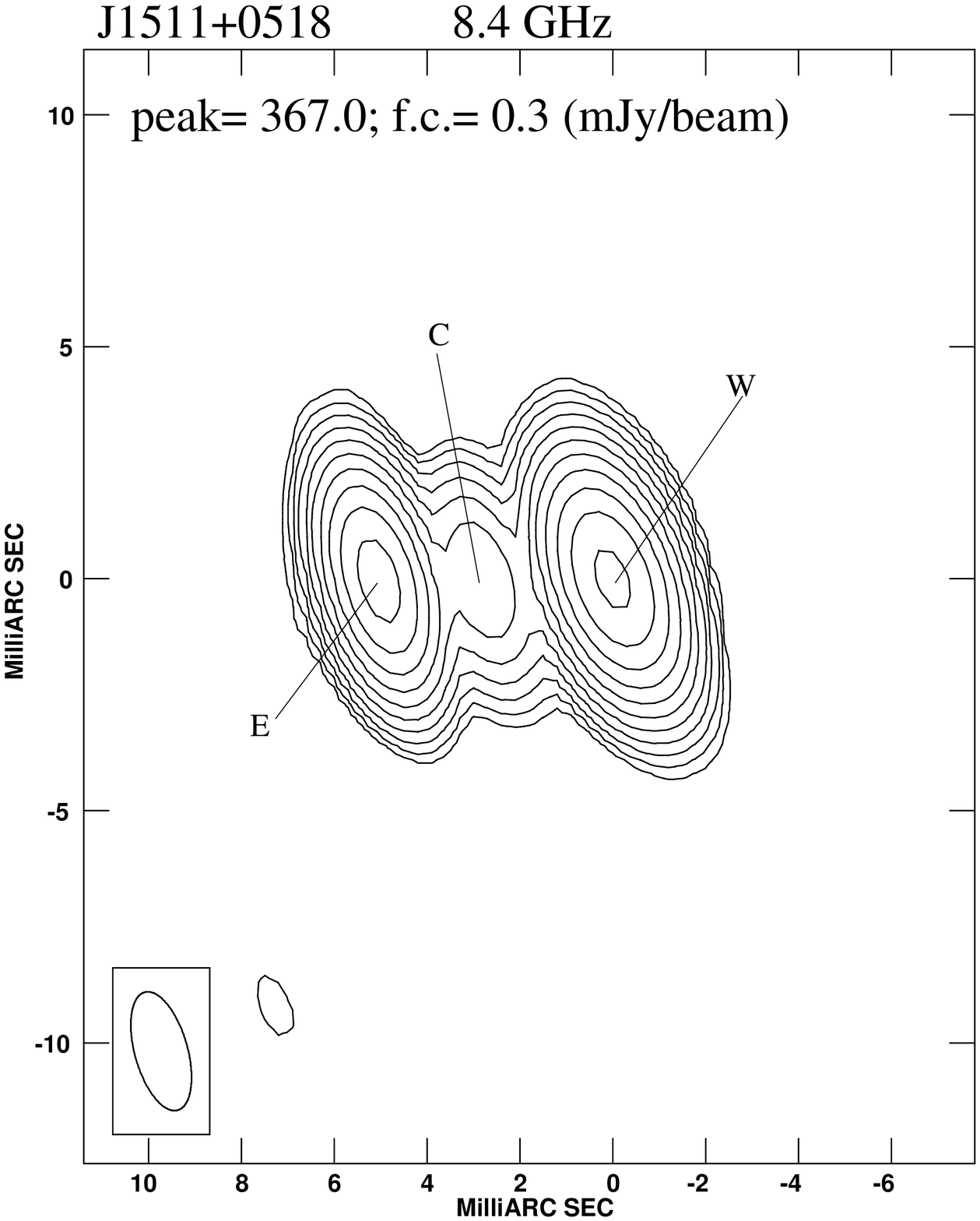}
\includegraphics{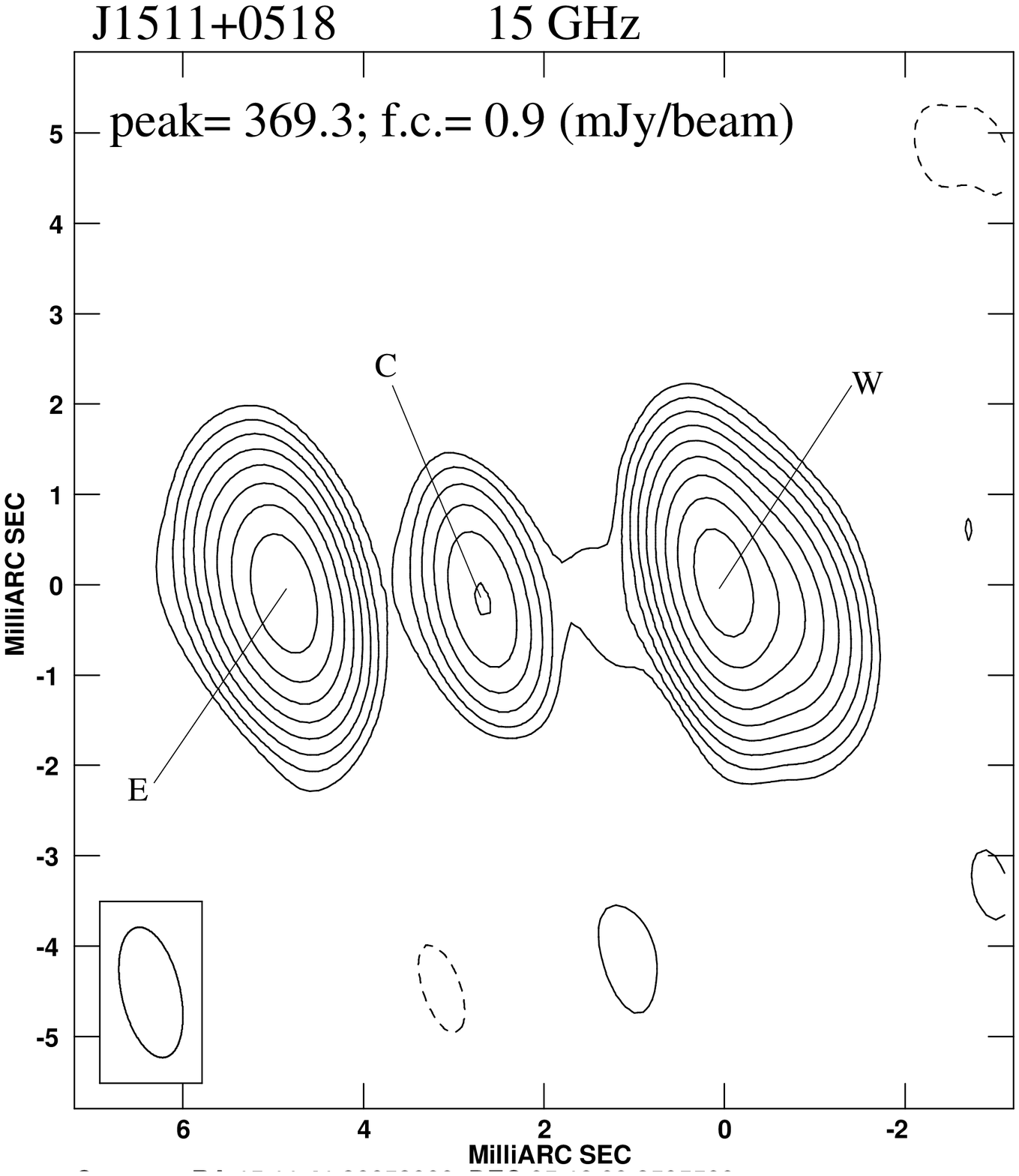}
\vspace{14cm}
\caption{VLBA images of the source J1511+0518 in L, S, C, X and U
  bands. VLBA observations were carried out on July 21 2006.
For each image, we provide the
  following information on the plot itself: {\bf a)} the source name
  and the observing frequency on the top left corner; {\bf b)} the peak
flux density in mJy/beam; {\bf c)} the first contour intensity ({\it
  f.c.} in mJy/beam), which is usually 3 times the off-source rms
noise level measured on the image; contour levels increase by a factor
2; {\bf d)} the restoring beam, plotted on the bottom left corner.}
\label{fig_1511}
\end{center}
\end{figure*}

\subsection{Magnetic fields}

In the literature there are a few estimates of the magnetic field
based on the determination of the spectral peak parameters. In general,
difficulties in measuring the component angular size and peak
frequency
with good precision
meant that it was not possible to derive 
accurate measurements of the magnetic field. 
The observing frequencies were not sufficiently close to the turnover
frequency of the source components, leading to uncertainties in the
determination of the magnetic field of a factor of ten (e.g. Rendong et
al. \cite{rendong91}).
Furthermore, complex source morphology and unsatisfactory {\it uv-}
and frequency coverages make this task extremely difficult.
With its high spatial resolution, 
the VLBA is one of the best instrument to perform such an investigation,
but nonetheless it has not been used for this purpose.\\   
Since both the overall radio spectra of HFPs and the spectra of their
single sub-components peak in the range of frequencies
well sampled by the VLBA, we can constrain with good accuracy the
turnover frequency, flux density, and angular size, and therefore achieve
an accurate estimate of the magnetic field.

\begin{table}
\begin{center}
\caption{Physical parameters of the source components. }
\begin{tabular}{c|c|c|c|c|c}
\hline
Source&C&$H$&$H_{eq}$&$u_{\rm min }$&$p_{\rm min}$\\
 & &mG&mG&erg/cm$^{3}$&dyne/cm$^{2}$\\
 & &  &  &10$^{-4}$&10$^{-4}$\\ 
\hline
&&&&&\\
J0003+2129&E &33&30&5.0&3.1\\
J0005+0524&E &-&18&0.75&0.46\\
J0428+3259&E &$\gg$1000&34&0.75&0.46\\
          &Ce&59&65&3.9&2.4\\
J0650+6001&N &29&77&6.0&4.0\\
          &S &10&54&1.5&1.0\\
J1511+0518&W &104&95&8.3&5.2\\
          &E &$\gg$1000&70&3.8&2.4\\
J1459+3337&  &160&160&24&15\\
&&&&&\\
\hline
\end{tabular}
\label{tab_mag}
\end{center}
\end{table}

Following the approach by Readhead (\cite{read94})
in Eq. \ref{h_synchro} we consider component angular sizes
that are 1.8 times larger than the full width at half maximum derived
by the Gaussian fit. The uncertainty in the magnetic field measurement
depends strongly
on the accuracy of source parameters, in particular 
the angular size and peak frequency.  \\
In general, we find that the magnetic field strengths derived from the observed
spectral peak parameters are in the range of $\sim$ 10-100 mG for the
various components of the target sources discussed here, with typical
uncertainties of a factor of 2 or even less. Two
remarkable exceptions are J0428+3259 East and J1511+0518 East, for
which we measured
a magnetic field strength that was higher than a few Gauss.
For these components, the radio peak 
appears to be inconsistent with pure synchrotron self-absorption and that a
magnetic field derived by adopting this assumption is unreliable 
(see Sect. 5.1 for a proper
discussion). In the case of J1511+0518 East, this interpretation is
supported by an optically-thick spectral index $< - 2.5$, which is the
lowest value achievable by SSA.
In both components, FFA is detected against the faintest 
  component. It is unclear whether this is due only
  to the presence of an additional absorber,
  which reduces the flux density.\\
We then derived the magnetic field by assuming that the source
components were in equipartition. Equipartition
magnetic fields were computed with standard formulae (Pacholczyk
\cite{pacho70}), assuming an ellipsoidal geometry with a filling factor
of unity (i.e. the source volume is fully and homogeneously filled with
relativistic plasma). Furthermore, proton and electron energies were
assumed to be equal. Equipartition values are
accurately determined, and the field computed in this way was relatively
insensitive to measured quantities that were poorly constrained by physical
parameters. The equipartition magnetic fields obtained are reported in
Table \ref{tab_mag} with physical parameters such as the
internal pressure and the energy density computed assuming a minimum
energy condition (Pacholczyk \cite{pacho70}).\\

\subsection{Hot-spot separation velocity}
 
For the 5 HFPs presented here, 
the availability of VLBA observations carried out
in 2002 enables us to estimate the hot-spot separation velocity.\\
The velocity determined in this way indicates the rate at
which the working surfaces of the jets in the ISM 
are increasing their separation.\\
For the source J0428+3259, the uncertainties in the position of the
components in the 2002 data 
are rather large and similar to the variation measured with respect
to the 2006 data, preventing a significant estimate of the source
growth based on these datasets only.\\
In the case of J1511+0518 where a central component is
present, we computed the separation of the outer components with
respect to the central one. The determination of 
the central component position in 2002 data was rather inaccurate
(0.1 mas, i.e. the pixel size), 
and thus we prefer to derive only the hot-spot separation velocity.\\
We note that we measure separation velocities between two
epochs only over a small time baseline, 
and the uncertainties in the estimated
values are therefore high. 
When only data at two epochs are available, the errors  
cannot be estimated by regression analysis, and they are
determined on the basis of the accuracy to which
component positions can be measured (Polatidis \& Conway
\cite{pc03}).\\
For the sources J0003+2129 and J1511+0518, we found 
hot-spot separation velocities
of 0.15$c$ and 0.1$c$, respectively, 
similar to those obtained for the same class of
objects, which are usually between 0.1$c$ and 0.4$c$.
(Polatidis \&
Conway \cite{pc03}). In the case of J0005+0524,
we measured a far higher velocity of $\sim$ 0.7$\pm$0.1$c$,
while in J0650+6001 no hot-spot separation was detected, in agreement
with previous works (e.g. Akujor et al. \cite{akujor96}). Among the young
sources, this is the only object that does not show any evidence of
proper motion so far.\\
We estimated the kinematic source age by assuming that the sources 
grow with a constant hot-spot separation velocity. Table
\ref{expansion} reports the hot-spot separation velocity, the total linear
size (LLS), and the kinematic age of the observed sources.\\

\begin{table}
\begin{center}
\caption{Hot-spot separation velocity and the total linear size LLS.}
\begin{tabular}{c|c|c|c}
\hline
Source&v&LLS&age\\
 &$c$&pc&yr\\
\hline
&&&\\
J0003+2129&0.15$\pm$0.10&21&500\\
J0005+0524&0.7$\pm$0.1&15&140\\
J0428+3259& - &16& - \\
J0650+6005& - &17& - \\
J1511+0518&0.10$\pm$0.01&7&250\\
&&&\\
\hline
\end{tabular}
\label{expansion}
\end{center}
\end{table}

\section{Discussion}

The direct measurement of the magnetic field in extragalactic radio sources
is a difficult task to carry out. An indirect way to estimate the
magnetic field is to assume that the radio source is in equipartition.
This assumption was supported by 
equipartition brightness temperatures, which were found to agree with 
maximum brightness temperatures (Readhead \cite{read94}).\\
Alternatively, the magnetic field can be determined by comparing
synchrotron and inverse Compton losses, but X-ray observations of
small and young radio sources have not provided strong constraints so
far, mainly due to the poorer spatial resolution of the X-ray
telescopes with respect to that achieved by radio interferometry.\\
Another way to test whether radio sources are in equipartition is
to compute their magnetic fields by means of observable quantities. In
sources showing SSA, the value of the field strength
can be 
derived directly from the synchrotron peak parameters. \\
In the following discussion, we investigate the physical conditions
occurring in young radio sources by comparing the results obtained by
using the peak parameters and those assuming equipartition conditions.\\

\subsection{Magnetic fields and the origin of the turnover}

For each source component, the magnetic fields that  
we estimated from the spectral peak determined from
the data (i.e. frequency and flux density) range between 10 and 100
mG, in agreement with values found in case of equipartition, with the
exception of J0428+3259 East and J1511+0518 East.
Such a result suggests 
that young radio
sources are in equipartition and the peak in their spectra
is probably due to synchrotron self-absorption.
Only in J0428+3259 East and J1511+0518 East, a field higher than a few
Gauss clearly exceeds the equipartition value.
In the presence of such a
high magnetic field, the radiative lifetime $t_{\rm rad}$
of the electron population should be extremely short:\\

\begin{equation}
t_{\rm rad} \sim 0.625 \, H^{-3/2} \nu_{\rm br}^{-1/2}\;\;\;\;({\rm yr})
\label{eq_rad} 
\end{equation}

\begin{figure}
\begin{center}
\includegraphics{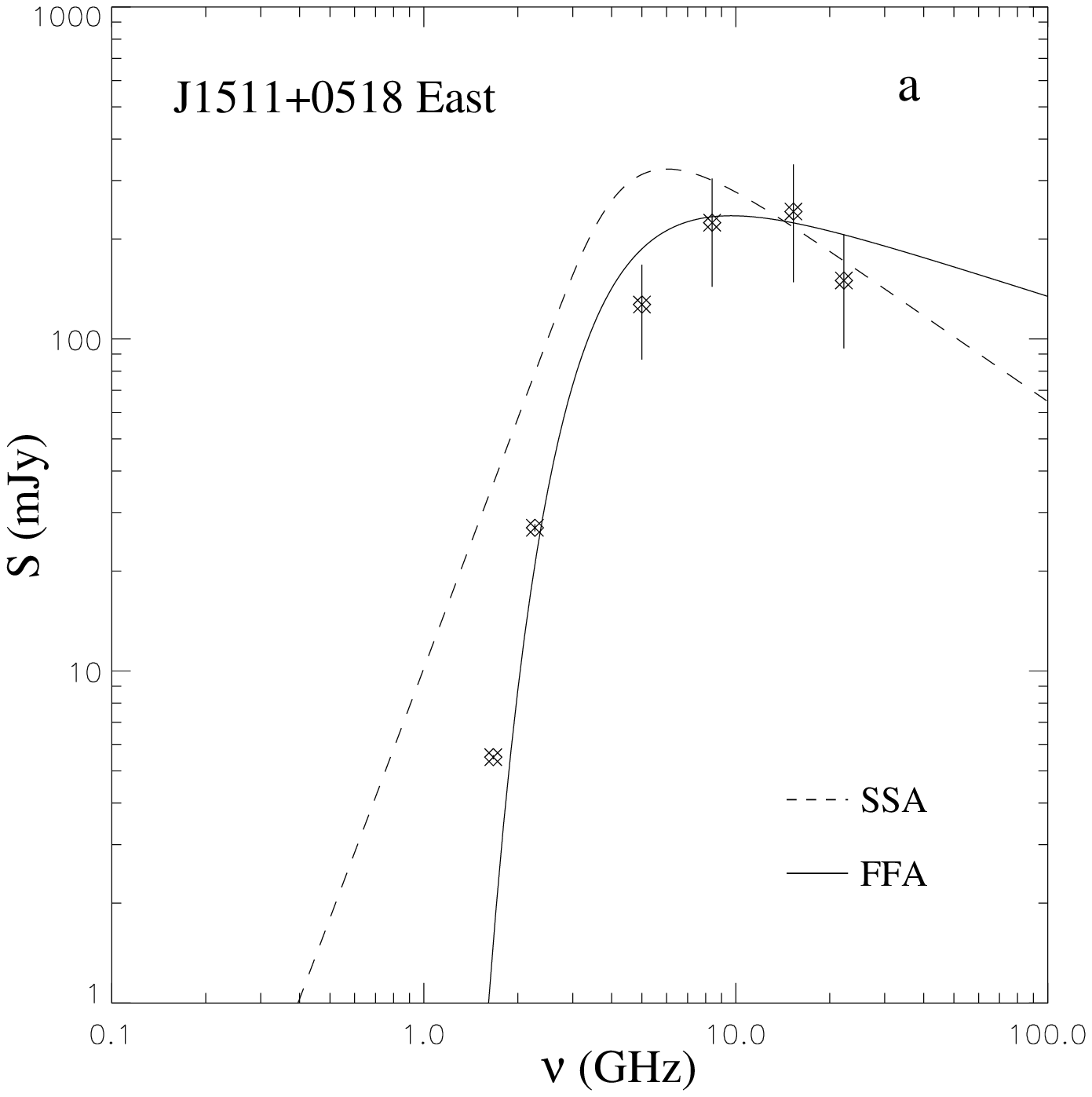}
\includegraphics{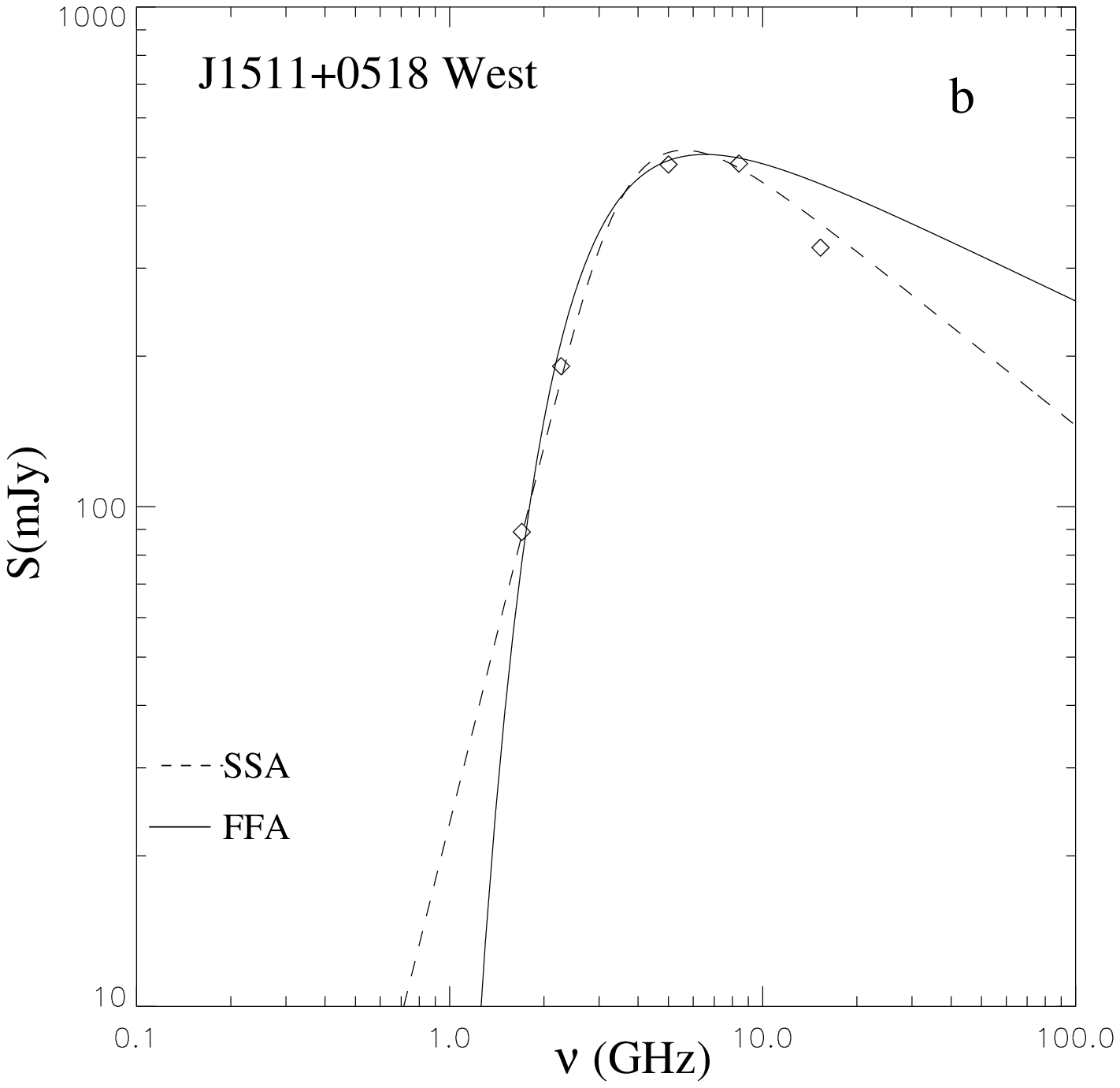}
\includegraphics{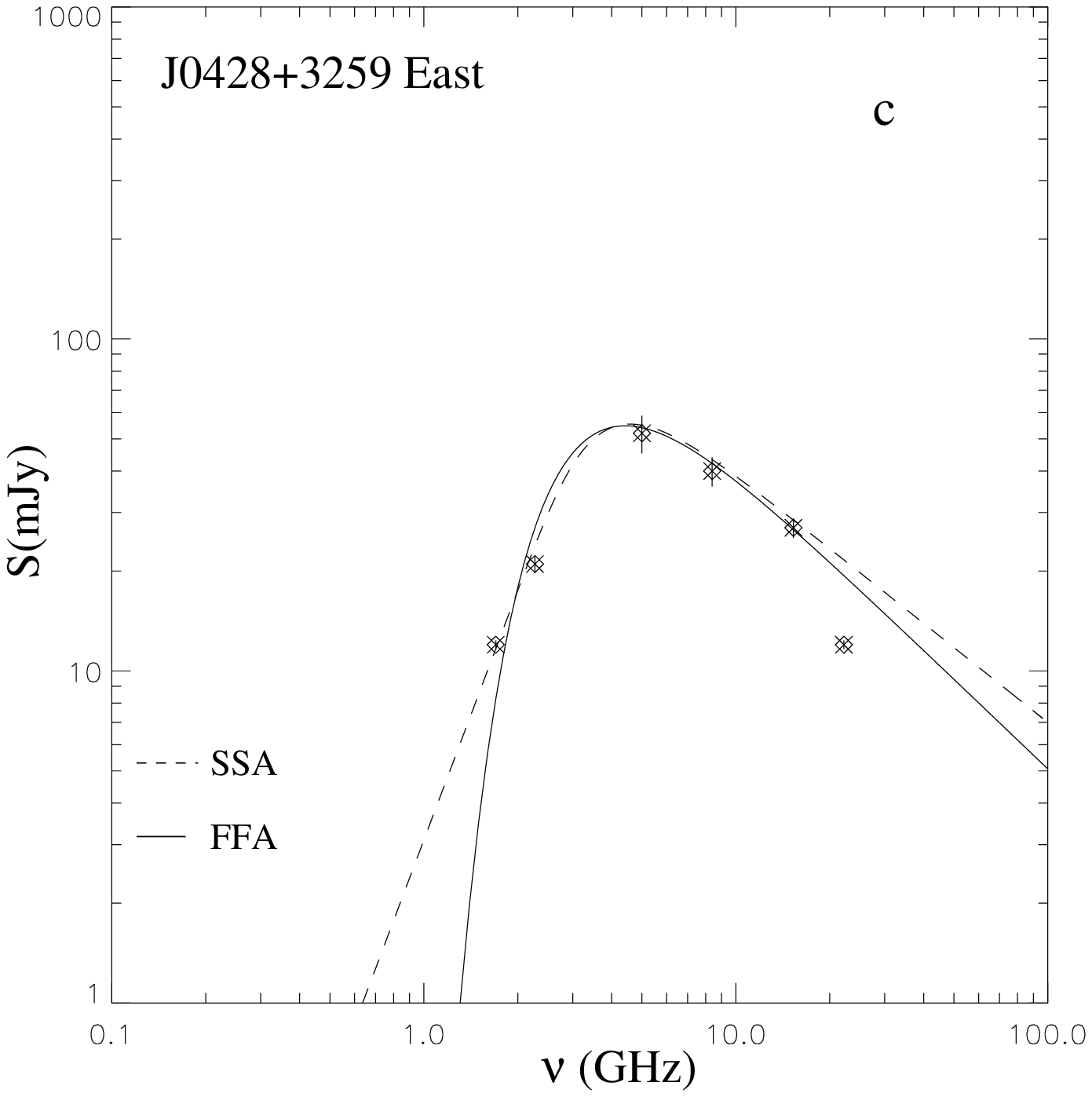}
\vspace{18cm}
\caption{FFA and SSA fit to the East ({\it top}) and West ({\it middle})
  components of J1511+0518 and to the East component of J0428+3259
  ({\it bottom}).}
\label{ffa}
\end{center}
\end{figure}

\noindent where $\nu_{\rm br}$ is the break frequency in GHz and $H$ in
Gauss. The radio spectrum of J1511+0518 East does not show any evidence
of the spectral break in the frequency range spanned by
  the observations. If we assume a lower limit of 22 GHz for the break
  frequency, we find that, in the presence of such high magnetic fields
  the electron population would have synchrotron ages 
$t_{\rm rad} \sim 1 {\rm yr}$. However, 
the spectral
coverage in the optically thin regime is rather sparse,
preventing a proper detection of the break.\\
A more realistic explanation is that the spectral peak is
due to
free-free absorption. To test the
reliability of this hypothesis, we fit the spectrum 
with both FFA and SSA. 
From Fig. \ref{ffa}a, it is clear that the spectrum is described more
  accurately by a
FFA model (solid line) than a SSA one (dashed line). 
This result, together with an optically-thick spectral index $<\; -2.5$,  
implies that the spectrum would be more accurately described by  
absorption by thermal plasma in front of the component, instead of
simple synchrotron self-absorption alone. 
On the other hand, the spectrum of the the western component of
J1511+0518 may be reproduced well by a pure SSA model without any
additional contribution from FFA.
The good agreement between the
magnetic field obtained by the peak
parameters and the equipartition value
suggests that in this case the spectral peak is probably due
to SSA. This result may indicate that the thermal plasma causing
the absorption in the East component does not embed homogeneously
the radio source, as in the presence of a diffuse and isotropic 
circumnuclear structure, but
is probably due to a inhomogeneous ambient medium of different 
opacity along the various lines of sight, as also suggested by Kameno et
al. (\cite{kameno00}) in the case of
the other HFP galaxy J1407+2827 (OQ\,208).\\
The same approach was carried out to model the East component of
J0428+3259 (Fig \ref{ffa}c). 
The magnetic field inferred from the peak parameters was 
$H \sim$ 10 G, which is, again, substantially higher than for the 
components of the other target sources. Even for this
component it is therefore likely that the spectral peak is due to absorption
from an ionized ambient medium in addition to synchrotron
self-absorption. \\
The discovery of free-free absorption in two of the five sources 
observed here may
indicate a detection rate of such an effect that is higher in HFPs than in
the larger GPS sources. The statistic is still limited and further
observations are required, but this evidence is consistent with the idea
that the smallest sources (i.e. HFPs) reside within the innermost
region of the host galaxy, characterized by an extremely dense and
inhomogeneous ambient medium of high electron density.
Other detections of free-free absorption occurring in  
the HFP galaxies J0111+3906 (Marr et al. \cite{marr01}) and J1407+2827
(alias OQ\,208, Kameno et al. \cite{kameno00}) support this view.\\ 
In such a context, we cannot exclude that these
components are in equipartition since we do not have
another way to derive independently the magnetic field. Ideally, it may be
measured by comparing synchrotron and inverse Compton losses, but 
X-ray observations available cannot provide a comparable spatial
resolution and the magnetic field derived would be averaged over the
entire source volume and not over individual components.\\
However, to estimate the optical depth $\tau$ of the ionized medium
we compute the peak frequency of these two components 
by assuming equipartition, since the
other targets do not show any significant departure from minimum
energy conditions. We find $\nu_{p} \sim$ 1.9 and 5.0 GHz for
J0428+3259 East and J1511+0518 East respectively, in the case of SSA only. 
Then we determine
the optical depth $\tau$ by comparing the observed flux density
$S_{\rm FFA}$
with that obtained by extrapolating the spectral index of the
optically-thin spectrum down to the peak ($S_{\rm SSA}$):\\

\begin{equation}
\tau = - {\rm ln} \frac{S_{\rm FFA}}{S_{\rm SSA}}
\end{equation}

\noindent and the values derived (Table \ref{tab_ffa})
are similar to those found in the
components of the young GPS radio source B1946+708 (Peck et al. \cite{ap99}).\\

\begin{table}
\begin{center}
\caption{Free-free optical depth of J0428+3259 East and
  J1511+0518. The optical  depth $\tau$ has been computed from Eq. 3.}
\begin{tabular}{c|c|c|c|c|c}
\hline
Source&$\alpha_{\rm thin}$&$\nu$&$S_{\rm FFA}$&$S_{\rm SSA}$&$\tau$\\
 & &GHz&mJy&mJy& \\
\hline
&&&&&\\
J0428+3259 E&0.6&2.27&21&84&1.3\\
            &0.6&5.0&50&55&0.09\\
J1511+0518 E &0.8&5.0&127&515&1.4\\
             &0.8&8.4&224&340&0.4\\
&&&&&\\
\hline
\end{tabular}
\label{tab_ffa}
\end{center}
\end{table}

\subsection{Magnetic field and the electron population}

As shown in the previous section, the magnetic fields obtained following
equipartition and, when possible, 
confirmed by the value inferred from the peak parameters are
in the range of $\sim$ 10 mG up to $\sim$160 mG, if we also consider 
GPS J1459+3337 (Table \ref{tab_mag}) 
With such strong fields, electron populations with relatively small
$\gamma$ can radiate at high frequencies.
The critical frequency $\nu_{cr}$ at which an electron 
with a certain $\gamma$ radiate is given by:

\begin{equation}
\nu_{cr} \sim 4.2 \times 10^{-3}  \gamma^{2} H\,\,\,({\rm GHz}) 
\label{eq_nucr}
\end{equation}

\noindent (Pacholczyk \cite{pacho70}), where $H$ is in Gauss.
From Eq. \ref{eq_nucr}, it is clear that
electrons with small $\gamma$ ($\sim 200 - 500$) can radiate at
frequencies higher than a few GHz in the presence of $H$ in the range of
10-100 mG. \\

\begin{figure}
\begin{center}
\includegraphics{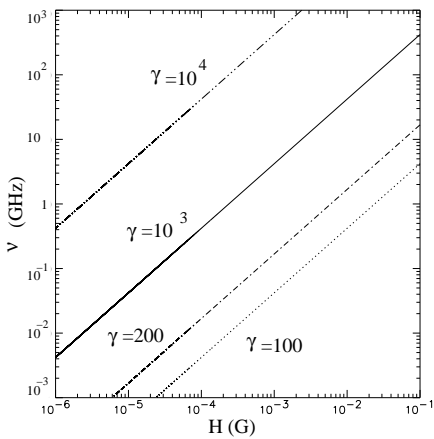}
\vspace{6cm}
\caption{Critical frequency emitted by electron populations with
  various $\gamma$ versus the magnetic field. $\gamma$ has been
  computed from Eq. 4.}
\label{fig:h_nu}
\end{center}
\end{figure}

If we assume that the jet power output $P_{\rm j}$ is constant during the lifetime of the
radio source and that the magnetic field is frozen into the plasma, we
find that the magnetic field intensity decreases as the source grows:\\

\begin{equation}
H = \left( \frac{P_{\rm j} t}{V} \right)^{1/2}
\label{eq_frozen}
\end{equation}

\noindent where $t$ is the source age and $V$ is the volume of the lobe
	  computed by assuming an ellipsoidal geometry and a filling
	  factor of unity.\\
The decrement of the magnetic field as the source grows in size is in 
agreement with the observations: compact (LLS $\ll$ 20 kpc) objects
have magnetic fields of a few mG (e.g. Dallacasa et al. \cite{dd02};
Fanti et al. \cite{cf95}),
while in large (LLS $\sim$ 1 Mpc) radio sources Murgia et
al. (\cite{mm01}) found a magnetic field of a few $\mu$G.\\
From Eq. \ref{eq_nucr}, we see that the critical frequency at which an
electron with a given $\gamma$ radiates, decreases with
the magnetic field (Fig. \ref{fig:h_nu}). For example, 
in the presence of high magnetic fields ($H \geq 10$ mG), 
such as those found in compact objects, an electron with
$\gamma \sim 1000$ emits at a frequency around 40 GHz, while in
weaker fields ($H \sim 2 \mu{\rm G}$) the radiation is at a 
frequency about
of $\sim$ 10 MHz, which is lower than the frequencies of the surveys
carried out so far. 
As a consequence, during the evolution of a radio source the field
intensity progressively decreases and the electrons radiating at a
given frequency are those with higher and higher $\gamma$. For example,
electrons with $\gamma \sim 350$ emit at 5 GHz in the presence of $H = 10$
mG, while for $H = 2 \mu$G only electrons with $\gamma \sim
24000$ are still visible at such a frequency.  
It is unclear whether all the small sources may become efficient
accelerators to achieve $\gamma \sim 10^{3-4}$ as observed in extended
radio sources. On the other hand, electrons with such energies, in the
  presence of a magnetic field $H \sim 0.1$ G,
  radiate at $\nu_{\rm cr} \sim 2.5 \times 10^{14} {\rm Hz}$, i.e. in
  the near-IR. Their contribution to the NIR emission is however 
  limited, well below 1 mJy, as can be seen easily by extrapolating
  the optically thin spectra for all the sources presented
  here. At such a frequency and with such a high magnetic
  field, their radiative lifetime is extremely short and any
  synchrotron emission would be strongly contaminated by the thermal
  emission from the host galaxy.
All this may provide further explanation for the open
question concerning the high number counts of young radio sources with
respect to the larger ones, if we assume that some of young radio
sources will never be able to accelerate electrons to $\gamma > 10^{3}$.\\

\subsection{Extended emission and discontinuous radio activity}

Another intriguing aspect of these young radio sources is the
presence of low-surface brightness extended emission visible at low
frequencies on the pc-scale. In the source J1511+0518, 
we detected low-surface
brightness emission located about 50 pc from the main source at a position
angle of 110$^{\circ}$ (Fig. \ref{fig_1511}).
Extended emission on the pc-scale was discovered 
in the HFP galaxy J1407+2827 (Luo et al. \cite{luo07}). 
Extended emission on the kpc-scale and beyond was already known in
this class of objects (e.g. the HFP galaxy J0111+3906, Baum
\cite{baum90}), and interpreted in terms of the relic of
a past radio activity which occurred about 10$^{7}$ - 10$^{8}$ years
ago. 
Extended features located at a pc-scale distance from the central
object may be the
relic of a far more recent previous 
activity that occurred about $10^{3} -
10^{4}$ years ago.\\ 
Evidence of a ``recent'' earlier  
start may suggest that at the very beginning of the radio activity
several cycles of subsequent short bursts (sputtering)
occurred before 
the classical large Double radio sources started to develop. 
This was also suggested by Gugliucci et
al. (\cite{gugliu05}), who 
estimated source ages of a few
hundred years for the CSOs studied. The age distribution found peaks
around 500 years, suggesting that either many CSOs die young or that they are a
transient phenomenon and only a few become large Doubles
(Gugliucci et al. \cite{gugliu05}).   
Such ages agree with those derived
from the hot-spot separation velocities measured in our
sources. However, our values are affected by
large uncertainties: a more robust estimate of the velocity separation
and the kinematic age
would need more observing epochs distributed over several
years. \\

\section{Conclusions}

We have presented the results of multi-frequency VLBA observations
of 5 young High Frequency Peakers. By assuming that the turnover of the
radio spectrum is due to synchrotron self-absorption, 
we derived the magnetic field of each source sub-component by means
of observable quantities, namely the peak frequency, peak
flux-density, and the angular size.  
The magnetic fields calculated in this way are
usually in good agreement with those obtained by assuming equipartition,
with two exceptions. This suggests that in general
young radio sources are in minimum energy conditions and their
spectral turnover is caused by SSA. The values found for the magnetic
fields are in the range of 10 mG up to $\sim$ 160 mG, if we consider
also the GPS source J1459+3337.\\
In two sub-components, we found
evidence that the turnover of the spectrum is probably due to
absorption by a thermal plasma located in front of the component, instead
of SSA. 
The other components of these sources do not
show any evidence of departure from SSA, suggesting that the absorber is an
inhomogeneous ambient medium with different free-free opacities along
the lines of sight to the two lobes. \\
In young radio sources, the presence of such high magnetic fields 
implies that even electron populations with small $\gamma$ (e.g. $\gamma
\sim 200$) radiate at the ``high'' frequencies sampled by the
available surveys. This may provide a further explanation of the high
number counts of young radio sources with respect to the larger
ones.\\
In HFP J1511+0518 at 1.7 GHz,  
we unambiguously detected a low-surface brightness extended
emission not aligned with the main source structure. Such a feature
may be the relic of a previous epoch of radio activity that occurred
not long ago ($< 10^{4}$ years). This may be consistent
with a scenario of a discontinuous start (sputtering) of the radio activity.\\
The analysis of the hot-spot separation velocity provided kinematic
ages of a few hundreds of years, confirming that the targets can
be considered young radio sources.\\

\begin{acknowledgements}
The VLBA is operated by the US National Radio Astronomy Observatory
which is a facility of the National Science Foundation operated under
a cooperative agreement by Associated University, Inc.
This work has made use of the
NASA/IPAC Extragalactic Database (NED), which is operated by the Jet
Propulsion Laboratory, California Institute of Technology, under
contract with the National Aeronautics and Space Administration.\\
\end{acknowledgements}

\end{document}